\documentclass[11pt,a4paper]{article}

\usepackage[colorlinks=true, linkcolor=black!50!blue, urlcolor=blue, citecolor=blue, anchorcolor=blue]{hyperref}
\usepackage[font=small,labelfont=bf,margin=0mm,labelsep=period,tableposition=top]{caption}
\usepackage[a4paper,top=3cm,bottom=2.5cm,left=2.5cm,right=2.5cm,bindingoffset=0mm]{geometry}

\usepackage{graphicx}
\usepackage{float}
\usepackage{afterpage}
\usepackage{epsfig,cite}
\usepackage{amssymb}
\usepackage{amsmath}
\usepackage{dsfont}
\usepackage{multirow}
\usepackage{url}
\usepackage{xcolor}
\usepackage{float}
\usepackage{afterpage}

\usepackage{url}
\usepackage{hyperref}

\usepackage{multirow,booktabs,multirow}

\usepackage{graphicx}
\usepackage{float}
\usepackage{afterpage}
\usepackage{epsfig,cite}
\usepackage{amssymb}
\usepackage{amsmath}
\usepackage{dsfont}
\usepackage{multirow}
\usepackage{url,hyperref}
\usepackage{booktabs}
\usepackage{csquotes}
\MakeOuterQuote{"}

\usepackage{url}
\usepackage{hyperref}

\newcommand{\be}{\begin{equation}}
\newcommand{\ee}{\end{equation}}
\newcommand{\bea}{\begin{eqnarray}}
\newcommand{\eea}{\end{eqnarray}}
\newcommand{\bi}{\begin{itemize}}
\newcommand{\ei}{\end{itemize}}
\newcommand{\ben}{\begin{enumerate}}
\newcommand{\een}{\end{enumerate}}

\newcommand{\lp}{\left(}
\newcommand{\rp}{\right)}
\newcommand{\as}{\alpha_s}

\newcommand{\amz}{\alpha_s\left( m_Z \right)}

\def\frac#1#2{{{#1}\over {#2}}}
\def\gsim{\mathrel{\rlap{\lower4pt\hbox{\hskip1pt$\sim$}}
    \raise1pt\hbox{$>$}}}         %greater than or approx. symbol
\def\lsim{\mathrel{\rlap{\lower4pt\hbox{\hskip1pt$\sim$}}
    \raise1pt\hbox{$<$}}}         %less than or approx. symbol

\newcommand{\draft}[1]{}

\def\beq{\begin{equation}}  
\def\eeq{\end{equation}}  

\def \n0{N_j^{(0)}}

\def\lapprox{\lower .7ex\hbox{$\;\stackrel{\textstyle <}{\sim}\;$}}
\def\gapprox{\lower .7ex\hbox{$\;\stackrel{\textstyle >}{\sim}\;$}}

\numberwithin{equation}{section}
\numberwithin{figure}{section}
\numberwithin{table}{section}

\begin{document}
\newgeometry{top=1.5cm,bottom=1.5cm,left=2.5cm,right=2.5cm,bindingoffset=0mm}
%%%%%%%%%%%%%%%%%%%%%%%%%%%%%%%%%%%%%%%%%%%%%%%%
\begin{figure}[h]
  \includegraphics[width=0.32\textwidth]{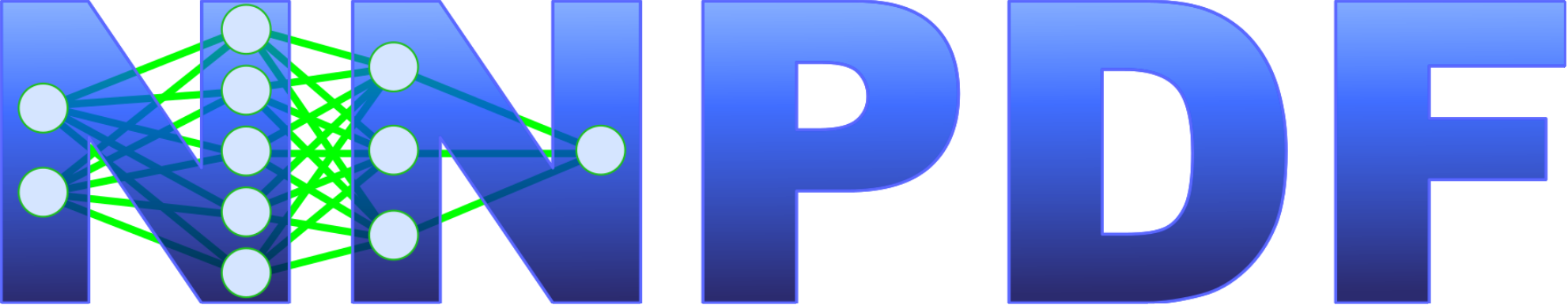}
\end{figure}
%%%%%%%%%%%%%%%%%%%%%%%%%%%%%%%%%%%%%%%%%%%%%%%%%
\vspace{-2.5cm}
\begin{flushright}
CAVENDISH-HEP-18-02\\
CERN-TH-2018-024\\
Edinburgh 2017/19\\
Nikhef/2017-046\\
OUTP-17-14P\\
DAMTP-2018-3\\
TIF-UNIMI-2018-1
\end{flushright}
\vspace{1cm}

\begin{center}
  {\Large \bf Precision determination of the strong coupling constant\\[0.2cm]
  within a global PDF analysis}
  \vspace{1.5cm}

  {\small
  {\bf  The NNPDF Collaboration:} \\[0.2cm]
  Richard~D.~Ball,$^{1}$ 
  Stefano~Carrazza,$^{2}$
  Luigi~Del~Debbio,$^{1}$ 
  Stefano~Forte,$^3$\\
  Zahari~Kassabov,$^{4}$
  Juan~Rojo,$^{5}$
  Emma~Slade,$^{6}$
  and Maria~Ubiali$^{7}$}

\vspace{1.0cm}
{\it \small ~$^1$ The Higgs Centre for Theoretical Physics, University of Edinburgh,\\
  JCMB, KB, Mayfield Rd, Edinburgh EH9 3JZ, Scotland\\[0.1cm]
  ~$^2$ Theoretical Physics Department, CERN, CH-1211 Geneva, Switzerland\\[0.1cm]
  ~$^3$ Tif Lab, Dipartimento di Fisica, Universit\`a di Milano and\\
INFN, Sezione di Milano, Via Celoria 16, I-20133 Milano, Italy\\[0.1cm]
~$^4$  Cavendish Laboratory, University of Cambridge, Cambridge CB3
	0HE, United Kingdom\\[0.1cm]
~$^5$ Department of Physics and Astronomy, VU University, NL-1081 HV Amsterdam,\\
and Nikhef Theory Group, Science Park 105, 1098 XG Amsterdam, The Netherlands\\[0.1cm]
  ~$^6$ Rudolf Peierls Centre for Theoretical Physics, 1 Keble
Road,\\ University of Oxford, OX1 3NP Oxford, United Kingdom\\[0.1cm]
~$^7$ DAMTP, University of Cambridge, Wilberforce Road, \\ Cambridge, CB3 0WA, United Kingdom
}

\vspace{1.0cm}

{\bf \large Abstract}

\end{center}
We present a  determination of the
strong coupling constant $\as(m_Z)$
based on the NNPDF3.1 determination of parton
distributions, which for the first time  includes  constraints
from jet production, top-quark pair differential distributions, and
the $Z$ $p_T$ distributions using exact NNLO theory.
Our result
is based on a novel extension of the NNPDF
methodology --- the correlated replica method --- which allows for 
a simultaneous determination of $\alpha_s$ and the PDFs with all
correlations between them fully taken into account.
We study in detail all relevant sources of experimental,
methodological and theoretical uncertainty.
At NNLO we find 
$\as(m_Z) = 0.1185 \pm 0.0005^\text{(exp)}\pm 0.0001^\text{(meth)}$,
showing that methodological uncertainties are negligible.
We conservatively estimate 
the theoretical uncertainty due to missing higher
order  QCD corrections (N$^3$LO and beyond)
from half the shift between the NLO and NNLO
$\alpha_s$ values, finding $\Delta\alpha^{\rm th}_s =0.0011$.

\clearpage

\tableofcontents

\section{Introduction}
\label{sec:introduction}

The value of the strong coupling constant $\amz$ is
a dominant source of uncertainty in the computation of several LHC processes.
This uncertainty is  often combined with that on parton
distributions (PDFs), with which it is strongly correlated.
However,
while PDF uncertainties have reduced considerably over the years, as
it is clear for example by comparing the 2012~\cite{Botje:2011sn} and
2015~\cite{Butterworth:2015oua} PDF4LHC recommendations, the
uncertainty on the $\as$ PDG average~\cite{Patrignani:2016xqp}
remains substantially
unchanged since 2010~\cite{Nakamura:2010zzi}.
As a consequence, the uncertainty on
$\as$ is now the dominant source of uncertainty for 
several  Higgs boson production cross-sections~\cite{deFlorian:2016spz}.

Possibly the cleanest~\cite{Altarelli:2013bpa,Salam:2017qdl}
determinations of $\as$ come from processes that do not require
a knowledge of the PDFs, such as the global
electroweak fit~\cite{deBlas:2016ojx}. These
are free from the need to control all sources of bias which may
affect the PDF determination and contaminate the resulting
$\as$ value.
A determination of $\as$ jointly with the PDFs, however, has the
advantage that it is driven by the
combination of a large number of experimental measurements from several
different processes.
This is advantageous
because possible sources of uncertainties related to specific measurements,
either of theoretical or experimental origin, are mostly
uncorrelated amongst each other and will  average out to
some extent in the final $\as$ result.
In addition to the above, the simultaneous global fit of $\as$ and the PDFs is
likely to be more precise and possibly also more accurate 
than individual determinations based on pre-existing PDF sets, many of which have 
recently
appeared~\cite{Johnson:2017ttl,Andreev:2017vxu,Klijnsma:2017eqp,Aaboud:2017fml,Bouzid:2017uak,Chatrchyan:2013haa,Britzger:2017maj}. This is due to the fact that it fully exploits the information
contained in the global dataset while accounting for the correlation of
$\as$ with the underlying PDFs. 

Here we present a determination of $\as$ which exploits the most
recent PDFs obtained with the NNPDF methodology, namely
NNPDF3.1~\cite{Ball:2017nwa}. This updates a previous determination of
$\as$~\cite{Lionetti:2011pw,Ball:2011us} based on
NNPDF2.1~\cite{Ball:2011mu,Ball:2011uy}.
In comparison to this previous PDF set, NNPDF3.1 represents
a substantial improvement
both in terms of input dataset, theoretical calculations,
and fitting methodology.
Specifically, NNPDF3.1 is
the first PDF set to make such an extensive use of LHC data as to be
dominated by them. It is in fact the first global analysis to simultaneously use
differential top, inclusive jet, and $Z$ $p_T$ distribution data, all using exact
NNLO theory.
Indeed, typical PDF uncertainties are of order of one to three
percent in the data region for NNPDF3.1, about a factor two smaller
than they were for NNPDF2.1.

\begin{figure}[t]
\begin{center}
  \includegraphics[width=0.90\textwidth]{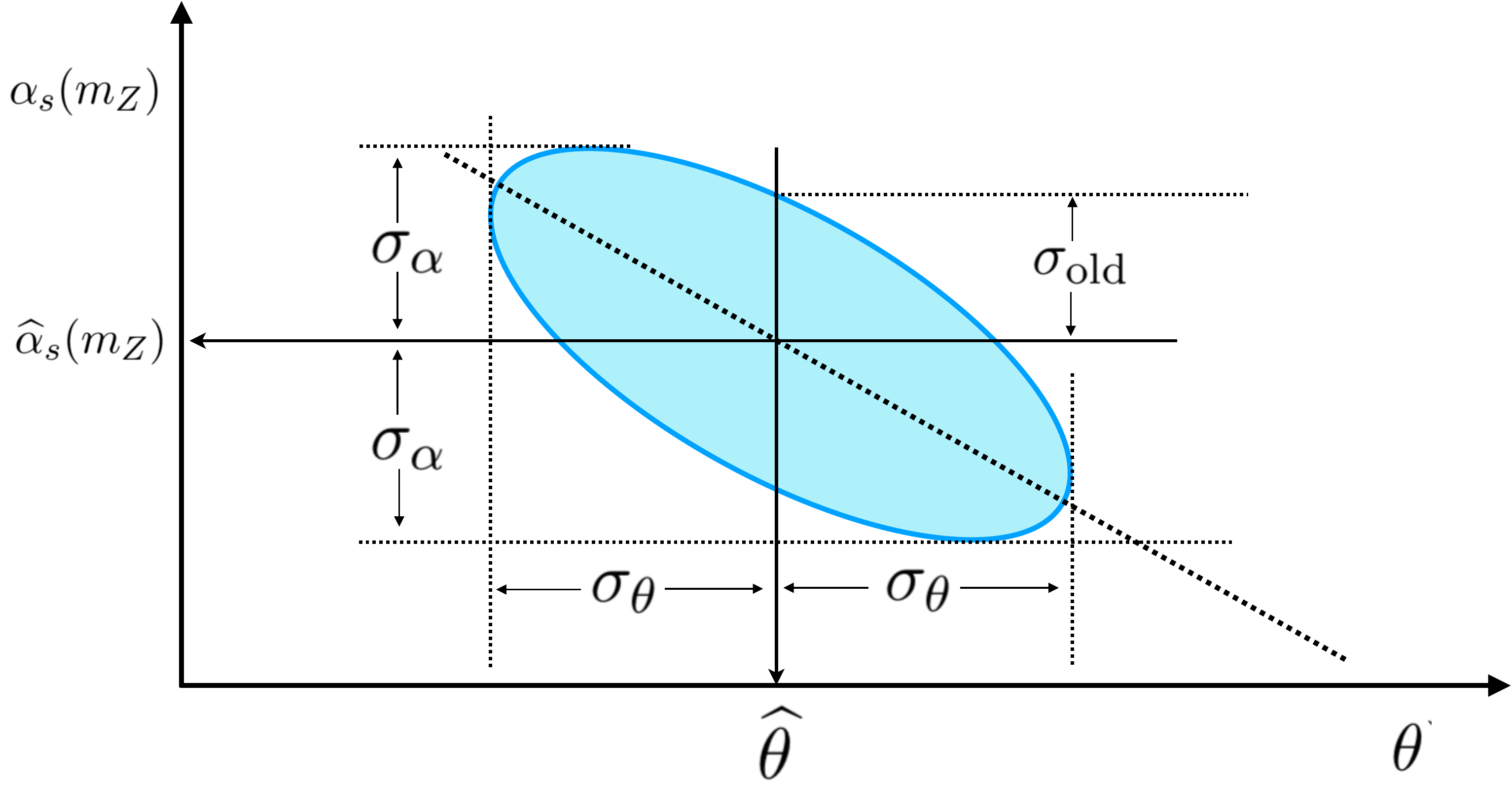}
  \caption{\small  Comparison between the standard deviation  of a pair of
correlated variables $(\as,\theta)$ and the one-sigma range for
the  variable $\as$  along
    the best-fit line of $\theta$. The best fit is denoted as $(\hat
    \as,\hat\theta)$ and the ellipse is the one-sigma
    contour about it. The standard deviations on  $(\as,\theta)$
    are $(\sigma_{\alpha},\sigma_\theta)$, while $\sigma_{\rm old}$ is
    the one-sigma interval for $\as$ with fixed $\theta=\hat\theta$.
}
    \label{fig:ellipses}
\end{center}
\end{figure}

This greater precision in the PDF determination requires
a corresponding improvement in the methodology used for the $\as$
extraction.
In our previous
work~\cite{Lionetti:2011pw,Ball:2011us}, PDF replicas were determined
for a number of fixed values of $\as$, which was then extracted
from the $\chi^2$ profile versus $\as$ of the best fit PDF,
obtained as an  average over the replicas.
Here instead,
both $\as$ and PDFs are determined from a simultaneous minimization
in their combined parameter space.
As we will discuss below, this
new method
corresponds roughly to determining the value and uncertainty on $\as$
from the error ellipse  of the multivariate measurement in
the $\lp \as, {\rm PDF}\rp$ hyperspace, and the old method corresponds
to performing a scan of  $\as$ along the best-fit
PDF line, see
 Fig.~\ref{fig:ellipses} for a schematic illustration.
In a situation when
the variables are highly correlated, especially if the
semi-axes of the ellipse are of very different length,  the
procedure used in our previous work might lead to an
underestimate of the uncertainty in $\as$. Hence the new procedure
becomes very relevant, now that some PDF
uncertainties are rather small.

%%%%%%%%%%%%%%%%%%%%%%%%%%%%%%%%%%%%%%%%%%%%%%%%%%%%%%%%%%%%%%%%%%%%%
%%%%%%%%%%%%%%%%%%%%%%%%%%%%%%%%%%%%%%%%%%%%%%%%%%%%%%%%%%%%%%%%%%%%%%

It turns out that the implementation of this simultaneous minimization
within the NNPDF methodology is nontrivial: it  requires the
development of a suitable generalization of the standard NNPDF
approach, which we call the correlated
replica method.
Using this strategy, $\as$ can be treated like any
other quantity that depends on the PDFs. In particular, its central value
and uncertainty can be determined
by performing statistics over a replica sample.
This means that, for example, the uncertainty on $\as$ is the standard deviation of an
ensemble of $\as$ values.
As we shall see, this allows for a
determination of $\as$ with small experimental uncertainties, and 
negligible methodological uncertainties.
Having reduced very much the size of all other uncertainties, the problem
of accurately estimating theoretical uncertainties becomes quite
serious. This is specifically problematic in the case of
missing higher-order uncertainties
(MHOUs), for which no fully satisfactory method has been
developed. Here we will conservatively estimate the theoretical uncertainty due to missing higher order QCD
corrections (N$^3$LO and beyond) from half the shift between the NLO
and NNLO $\alpha_s$ values.

This paper consists of two main parts.
First, in Sect.~\ref{sec:fitsettings} we present the correlated
replica method used
for the determination of $\as$, explain how it is used to estimate
the associated PDF uncertainties, and compare
it with the method used in previous NNPDF determinations.
Then, in Sect.~\ref{sec:results} we present our determination of $\as$
at  NLO and NNLO together with a careful assessment of all sources of
uncertainty.
Possible future developments are
briefly outlined in Sect.~\ref{sec:conclusion}.

\section{The correlated Monte Carlo replica method}
\label{sec:fitsettings}

As discussed in the introduction, the $\as$ determination
presented here 
differs from our previous
one~\cite{Lionetti:2011pw,Ball:2011us} because now the value of
$\as$ and its uncertainty are determined from a correlated fit
together with the  PDFs.
After briefly summarizing the
main aspects of the NNPDF methodology and the way it was used to
determine $\as$ in Ref.~\cite{Lionetti:2011pw,Ball:2011us}, we
describe the main idea of the new method, and then 
discuss the details of its implementation.
Only the salient
aspects of the NNPDF methodology will be recalled here; the reader is
referred to the original literature (see Ref.~\cite{Ball:2017nwa}, of
which we follow the notation,  and references
therein) and recent reviews~\cite{Forte:2010dt,Butterworth:2015oua,Gao:2017yyd} 
for a more detailed discussion.

\subsection{General strategy}
\label{sec:corrmc}

The NNPDF fitting methodology is based on constructing a Monte Carlo
representation of the original data sample consisting of pseudodata 
(Monte Carlo replicas of the original data), and fitting 
PDF replicas to these data replicas.
Specifically, starting with
an $N_{\rm dat}$-component vector of experimental points  $D$ with
components $D_i$, a set of
$N_\mathrm{rep}$  replicas $ D^{(k)}$ of the data is generated by means of:
\begin{equation}
    \label{eq:L2shift}
    D_i^{(k)} = \left(1 + r_i^{{\rm nor},k} \sigma_i^{\rm nor}\right)\,
    \Big(
      D_i + \sum_{p=1}^{N_{\rm sys}} r_{i,p}^{{\rm sys},k} \sigma^{\rm sys}_{i,p} + r_i^{{\rm stat},k} \sigma^{\rm stat}_i
    \Big)\, ,\quad  i=1,\dots,N_{\rm dat} \, ,
  \end{equation}
where $k=1,\ldots,N_{\rm rep}$; $\sigma_i^{\rm nor}$,  $\sigma_i^{\rm sys}$ and  $\sigma_i^{\rm stat}$ are normalization, systematic and
  statistical uncertainties, and $r_i$ are random numbers such that 
statistics over the replica sample reproduces the original statistical
properties of the data in the limit of large $N_\mathrm{rep}$.
For
example, this means that
\begin{equation}
\lim_{N_{\rm rep}\to\infty} \mathrm{cov}\left(D_i D_j\right)=C_{ij},
\label{covmat}
\end{equation}
where $\mathrm{cov}$ denotes the covariance over the replica sample
and $C_{ij}$ is the full experimental covariance matrix of the data.

A PDF replica  is then fitted to each data replica $ D^{(k)}$. In the
NNPDF approach, PDFs are parametrized using neural networks, in  turn
specified by a vector of parameters $\theta$.
In the most recent
NNPDF3.1 analysis, this vector $\theta$ has 296 components,
corresponding to 37 parameters for eight neural networks (for the up,
antiup, down, antidown, strange, antistrange, total charm and gluon PDFs).
Thus, for each data replica $ D^{(k)}$ a best-fit
$\theta^{(k)}$ is found by minimizing a figure of merit characterizing
the agreement between theory and data: 
\begin{equation}
  \label{eq:chi2not}
  \chi^2(\theta,D) = \frac{1}{N_{\rm dat}}
  \sum_{i,j} (T_{i}[\theta] - D_{i}) \, \lp C_{t_0}^{-1}\rp_{ij} \,
  (T_{j}[\theta] - D_{j} )\, .
\end{equation}
Here, $T_{i}[\theta]$ is the theoretical prediction for the $i$-th
datapoint, dependent on the set of parameters $\theta$, and 
$C_{t_0}$ is the covariance matrix used in the fit.
Recall that in the presence
of multiplicative uncertainties, $C_{t_0}$ cannot be directly identified with
the experimental covariance matrix $C$ used for pseudodata generation
Eq.~(\ref{eq:L2shift})
lest the fit be biased~\cite{D'Agostini:1993uj}, 
and must thus be constructed instead using a
suitable procedure such as the $t_0$ method~\cite{Ball:2009qv}
(see also~\cite{Ball:2012wy}).

A peculiarity of the NNPDF approach 
is that the best-fit parameters of each replica, $\theta^{(k)}$, are not defined as the
absolute minimum of the $\chi^2$ Eq.~(\ref{eq:chi2not}) in order to avoid
overfitting, i.e. in order not to fit
statistical fluctuations.
Instead, a suitable
cross-validation algorithm is employed~\cite{Ball:2014uwa}.
We thus obtain a set of best-fit PDF replicas
$\theta^{(k)}$, each determined as the minimum with respect to
$\theta$ of the figure of merit
${\chi^{2(k)}}$ computed using the $k$-th data replica:
\begin{align}
\theta^{(k)}&=\mathrm{argmin}\left[\chi^2(\theta,D^{(k)})\right]\, ,
\label{eq:thetakdef}
\end{align}
where $\mathrm{argmin}$ should be understood as minimization through
cross-validation, rather than as the absolute minimum.
Note that, because we employ non-deterministic minimization
algorithms, specifically genetic algorithms,
the best-fit $\theta^{(k)}$
corresponding to a given data replica $ D^{(k)}$ is not
unique; two identical data replicas  $ D^{(k_1)}=D^{(k_2)}$  may
lead to two different $\theta^{(k_1)}\not=\theta^{(k_2)}$ in two runs of
the minimization algorithm. 

In summary, the standard NNPDF methodology produces a set of replicas
 $ D^{(k)}$ of the original data, and uses them to construct a set of PDF
replicas which correspond to parameters $\theta^{(k)}$, where $k$
runs over the replica sample.

The theory predictions $T_i$,  which enter in the figure of merit of
the fit Eq.~(\ref{eq:chi2not}) depend not only on the PDF parameters
$\theta$, but also on theory parameters, specifically the value of
$\as$.
Therefore, in general we can view the figure of merit as a function
$\chi^2(\as,\theta,D)$.
In standard NNPDF determinations,
$\as$ is treated as a fixed parameter, along with all other
theory parameters, such as quark masses, CKM matrix elements, the
fine structure constant, and so on.
On the other hand, it is well known (see
e.g. Ref.~\cite{Martin:1995gx} for an early reference) that
PDFs are strongly correlated to the value
of $\as$, so a determination of the combined PDF+$\as$
uncertainty on a process which depends on both, requires knowledge of the
PDFs as $\as$ is varied.
With this motivation, NNPDF sets are routinely released
for different fixed values of $\as$,
where the procedure of generating
data replicas  $D^{(k)}$  and determining PDF
replicas determined by the best-fit parameters
$\theta^{(k)}$ is repeated several times for different values of $\as$.

In our previous work~\cite{Lionetti:2011pw,Ball:2011us}, $\as$
was determined by first producing PDF fits for a range of values of
$\as$. The $\chi^2(\as)$ of the mean of all the replicas
was then fitted to a parabola
 as a function of $\as$.
This methodology has two main drawbacks. The first is that, as
mentioned, the PDFs are strongly correlated to the value of
$\as$.
With this method, however, the $\chi^2$ profile is determined as
a function of $\as$ along the line in $\theta$ space which
corresponds to the best-fit $\theta$ at each particular value of
$\as$, without taking into account the variations in $\theta$
space.
Hence, as illustrated in Fig~\ref{fig:ellipses},
with the methodology of Refs.~\cite{Lionetti:2011pw,Ball:2011us} the resulting
uncertainty on $\as$ could be somewhat underestimated.

The second drawback is more subtle.
In the NNPDF procedure, the PDF
uncertainty is determined from statistics over the replica
sample, so a one-sigma interval is determined by computing a standard
deviation over replicas.
Whether or not this corresponds exactly to a one-sigma (i.e.
$\Delta \chi^2 =1$) interval in $\as$ space is unclear.
In fact,
in PDF determinations based on Hessian minimization in
parameter space, the $\Delta \chi^2 =1$ criterion is modified by a
suitable tolerance factor~\cite{Pumplin:2001ct,Martin:2002aw,Martin:2009iq}, which possibly
accounts for data inconsistencies or parametrization bias.
It is unclear, but certainly possible, that 
PDF uncertainties estimated in the NNPDF fits also include, at least to some extent, such a
tolerance.

Ideally, we would like a method of determining $\as$ in which
the uncertainty on $\as$
is determined on exactly the same footing as the PDF uncertainty, and 
which thus yields the full probability distribution for $\as$, 
marginalised with respect to the PDF parameters.
 The goal is to
treat
$\as$ on the same footing as the vector of
parameters $\theta$ that determine  the PDFs, i.e. to
simultaneously minimize the figure of merit with
respect
to both $\as$ and  $\theta$.
This is difficult in practice,
because the  dependence
on $\as$ appears in the theoretical
predictions, which, for reasons of computational efficiency, are
provided in the form of pre-computed grids determined
 before the fit using the {\tt APFELgrid}
framework~\cite{Bertone:2013vaa,Bertone:2016lga}.

This difficulty can be overcome through the correlated replica method,
as we now explain.
The method relies on the
concept of ``correlated replica'', or c-replica for short.
A c-replica is a correlated set of PDF replicas, all obtained by
determining the  best-fit $\theta^{(k)}$ Eq.~(\ref{eq:thetakdef})
but with different (fixed) values of $\as$: given the data replica
$D^{(k)}$,  the minimization Eq.~(\ref{eq:thetakdef}) is performed
several times, for a range of fixed values of $\as(m_Z)$.
 A c-replica thus
corresponds to as many standard NNPDF replicas as the number of values
of $\as$ for which the minimization has been performed, all
obtained by fitting to the same underlying data replica $D^{(k)}$.

One can then determine the best-fit value $\as^{(k)}$ for the
$k$-th c-replica by minimizing as a function of $\as$ 
the figure of merit $\chi^2$ Eq.~(\ref{eq:chi2not}) computed with
$\theta^{(k)}(\as)$ as $\as$ is varied for fixed $k$.
Namely, we first define the figure of merit computed for the $k$-th c-replica,
\begin{equation}
\chi^{2(k)}(\alpha_s)=\chi^{2} \lp\alpha_s, \theta^{(k)}(\alpha_s),D^{(k)} \rp\, ,
\label{eq:chi2kdef}\end{equation}
which we can view as a function of $\as$. Note that the dependence of the theory
prediction $T$ and thus of the figure 
of merit Eq.~(\ref{eq:chi2not}) on $\alpha_s$ is both explicit, and
implicit through the best-fit parameters $\theta^{(k)}(\alpha_s)$.
We then determine the best-fit value of $\as$ for the $k$-th
c-replica as
\begin{equation}
\as^{(k)}=\mathrm{argmin}\left[\chi^{2(k)}(\as)\right].
\label{eq:alphakdef}\end{equation}
Note that while,
as discussed above, in order to avoid overfitting,  
the best-fit $\theta^{(k)}$ is not the absolute minimum of the figure
of merit, no overfitting of $\as$ is possible, because overfitting
happens when fitting a function, not a single parameter.
Hence, in
Eq.~(\ref{eq:alphakdef}) the best fit value $\as^{(k)}$ does
denote the absolute minimum.
Therefore, in practice $\as^{(k)}$
can be determined by  fitting a parabola to the discrete set of values of
$\chi^2(\as)$ for each replica, and finding the  minimum of the parabola.

Note also that determining the best-fit for the $k$-th c-replica
by first minimizing with respect to $\theta$ and then minimizing with
respect to $\as$ is equivalent to simultaneously
minimizing in the $(\as,\theta)$ hyperspace, provided
the same figure of merit is used for PDF and $\as$
determination.
For instance, the absolute
minimum in $(\as,\theta)$ is the solution to the coupled equations
\begin{eqnarray}
    \frac{\partial }{\partial \theta}\chi^2(\as,\theta)
    &=& 0\, ,
\label{eq:jointprobf}\\ 
    \frac{\partial}{\partial \as}\chi^{2}(\as,\theta)
    &=& 0\, ,
  \label{eq:jointproba}
\end{eqnarray}
where Eq.~(\ref{eq:jointprobf}) is actually a system of $N_{\rm par}$
equations because $\theta$ is an $N_{\rm par}$-component vector and
the partial derivative is a gradient.
On the other hand,
this solution can also be found (compare Fig.~\ref{fig:ellipses})
by first finding the solution
$\theta(\as)$  to Eq.~(\ref{eq:jointprobf}), determining
$\chi^2(\as) = \chi^{2}(\as,\theta(\as) )$, and solving
\begin{equation}
  \label{eq:prob}
  \frac{d}{d\as}\chi^{2}(\as) =
\left(\frac{\partial}{\partial\as} + 
\frac{\partial \theta}{\partial\as}
\frac{\partial }{\partial \theta }
    \right)\chi^{2}(\as,\theta)=0 \, .
\end{equation}
This two stage procedure yields the same solution as the coupled 
Eqs.~(\ref{eq:jointprobf})--(\ref{eq:jointproba}) because  the second
term in brackets on the r.h.s. of Eq.~(\ref{eq:prob}) vanishes since 
$\theta(\as)$  was the solution of Eq.~(\ref{eq:jointprobf}).

One thus ends up, for each data replica $D^{(k)}$, with a best fit value
$(\as^{(k)},\theta^{(k)})$ of both $\as$ and the PDF
parameters.
That is, from each c-replica we extract a single
best fit value $\as^{(k)}$ --- an ``$\as$ replica'' --- exactly 
on the same footing as all the other fit parameters. 
The ensemble of values $\as^{(k)}$ obtained from
all the c-replicas then provides a representation
of the probability density of $\as$ 
from which we can perform  statistics in the usual way.
Interestingly, this means that we can now not only compute the
best fit $\as$ and its uncertainty as the mean and standard deviation
(or indeed 68\% confidence interval) using the $\as$ replicas, 
but also the correlation
between  $\as$  and the PDFs or indeed any PDF-dependent quantity.

In summary, the correlated replica method is akin to the 
standard NNPDF methodology in that it starts by  producing a set of replicas
of the original data, but uses these  to construct a set of 
correlated $\as$-dependent PDF replicas, the c-replicas, which 
correspond to parameters $\theta^{(k)}(\as)$ when $k$
runs over the replica sample and $\as$ takes a number of discrete
values. From each c-replica a best-fit $\as^{(k)}$ can then be
determined, so each c-replica yields an $\as$
replica, with $\as^{(k)}$ defined by Eq.~(\ref{eq:alphakdef}).

Hence, the correlated replica method exploits the fact that in the
NNPDF approach it is sufficient to know the best-fit set of parameters
for each replica, and all other information is contained in the
replica sample.
The price to pay for this is that
the statistics of the $\as$  fitting is 
inevitably more demanding than with the method of
Refs.~\cite{Lionetti:2011pw,Ball:2011us} 
because we have now have to fit a different parabola for each c-replica.
The issues
arising from this will be discussed in the next section.

\subsection{Implementation}
\label{sec:strategy}

Building on the conceptual strategy described above,
we now present the practical implementation
of the correlated replica method.
As already mentioned, the best-fit $\as^{(k)}$
Eq.~(\ref{eq:alphakdef}) for the $k$-th c-replica
is determined by fitting a parabola to the figure of merit
$\chi^2(\as)$,  viewed as a
function of $\as$, known  at the discrete
set of $\as$ values for which best-fit $\theta^{(k)}(\as)$
are available.
The reliability of the quadratic approximation to
$\chi^{2(k)}$ Eq.~(\ref{eq:chi2kdef})
  and the stability of the position of the minimum
upon inclusion of higher order terms can be studied using standard
methods and will be discussed in Sect.~\ref{sec:tests} below.

The best-fit $\as$ and its uncertainty are then
determined, according to standard NNPDF methodology, as the mean and
standard deviation  computed over the sample of $\as$ replicas
\begin{equation}
  \label{eq:ascv}
 \as= \langle\as^{(k)}\rangle_{\rm rep};\quad \sigma_\alpha
 =\mathrm{std} \left(\as^{(k)}\right)_{\rm rep},
\end{equation}
where $\as^{(k)}$ is given by
Eq.~(\ref{eq:alphakdef}). 

The uncertainty due to the finite size of the replica sample can be
estimated by bootstrapping.
To this purpose,  one constructs $N_{\rm res}$ resamples of the
original sample of  $N_{\rm rep}$ values $\as^{(k)}$.
Each resample is obtained by drawing at random
$N_{\rm rep}$ values from the original sample by allowing
repetition.
This means that each resample differs from the original
sample because some values are repeated and others are missing. The
finite-size uncertainty is then estimated by first computing the mean $\as^{({\rm res}, i)}$ for each of the resamples, 
\begin{equation}
\label{eq:bootstrapmean}
\as^{({\rm res}, i)}=\langle
\as\rangle_{\rm rep} \, ,
\end{equation}
where the mean is computed over the $N_{\rm rep}$ values of the
$i$-th resample.
The bootstrapping estimate of the finite-size
uncertainty on the central value of $\as$
is then the standard deviation of the set of $\as^{({\rm res}, i)}$
\begin{equation}
\label{eq:bootstrapmeansig}
\Delta_{\as} = 
\mathrm{std} \left( \as^{({\rm res}, i)}\right)_{\rm res}.
\end{equation}
The uncertainty on the uncertainty $\Delta_{\sigma}$
can be similarly computed by first
determining the uncertainty Eq.~(\ref{eq:ascv}) for each resample,
thus leading to an uncertainty $\sigma_\alpha^{({\rm res}, i)}$, and
then  computing the standard deviation of the ensuing uncertainties:
\begin{equation}
\label{eq:bootstrapstd}
\Delta_{\sigma} = 
\mathrm{std} \left( \sigma_\alpha^{({\rm res}, i)}\right)_{\rm res}.
\end{equation}
We find that  results become independent of the random seed used to
generate the bootstrapping resamples when $N_\mathrm{res}\simeq10000$.

It turns out that, when determining the best-fit $\theta^{(k)}(\as)$
through the standard NNPDF minimization algorithm, a certain amount of
fluctuation of individual values of $\chi^2(\as)$  about the
parabolic best-fit is observed.
In other words, the $\chi^2$ profiles as a function $\as$ are
not very smooth.
It is therefore advantageous to
introduce an improvement of the algorithm, called batch
minimization,
which increases its accuracy at the cost of increasing the time required for
fitting.

Furthermore, when using the standard NNPDF minimization,
occasionally the fit fails to satisfy a number of
convergence and quality criteria
(see Sect.~3.3.2 of Ref.~\cite{Ball:2014uwa}), in which
case  it is discarded.
Consequently, for some c-replicas
$\chi^2(\as)$ is not available for all $\as$ values.
One
must then decide on a sensible criterion for c-replica selection, with the most
restrictive criterion being to only accept c-replicas for which
all $\chi^2(\as)$ values are available, and the least restrictive one to
accept c-replicas for which at least three $\chi^2(\as)$ values
are
available so a parabola can be fitted.
We now discuss batch
minimization and replica selection criteria in turn.

The idea of batch minimization is to refit a given set of data
replicas more than once.
In order to improve
the smoothness of the  $\chi^2$ profiles obtained by the direct
use of NNPDF minimization, 
we exploit the fact that the minimization algorithm is not
deterministic, and thus simply rerunning the minimization from a
different random seed leads to a slightly different answer.
Each of these refits is called a batch.
For each
c-replica $k$ and each $\as$ value we then end up with several
best-fit results $\theta^{(k)}_i(\as)$, where $i$ runs over
batches.

We then pick for each c-replica $k$ and for each
$\as$ value  the batch which gives the best $\chi^2$.
We also impose the condition that at least two of the batches for the given
c-replica and $\as$ value have converged, in order to
mitigate the influence of outliers that narrowly pass the
post-selection fit criteria.
The dependence of results on the number of batches used can then be
assessed a posteriori by comparing results found with different numbers
of batches.

After batch minimization, we end up with a set of c-replicas
$\theta^{(k)}(\as)$ where, however, for several c-replicas, results
may be missing for one or more $\as$ values because convergence
was not achieved.
We must thus determine the minimum number of
$\as$ values $N_\mathrm{min}$ such that a c-replica is accepted.
The threshold $N_\mathrm{min}$ is chosen to ensure the stability of
results.
Curves with too few points lead to an
unreliable  parabolic fit, and thus an unreliable best-fit
$\as^{(k)}$ for that c-replica. This then leads to outlier values
of $\as^{(k)}$ and a spuriously large value of the uncertainty
on the  $\as^{(k)}$ determination.
On the other hand, once the number of points is sufficient for a
reliable parabolic fit, requiring more points does not improve the
determination of $\as^{(k)}$, but it reduces the number of
c-replicas which are retained in the final sample, which in turn
increases the finite-size uncertainty.

Therefore, the optimal value of $N_\mathrm{min}$ arises  from a trade-off between
the uncertainty on $\as^{(k)}$ from the parabolic fitting, and the
finite-size uncertainty.
In order to keep both criteria into account,
we fix  $N_\mathrm{min}$ by
minimizing
the bootstrapping uncertainty
$\Delta_{\sigma}$ Eq.~(\ref{eq:bootstrapstd}). However, in order to
make sure that the selection is not  too tight, we do
not minimize  $\Delta_{\sigma}$ itself. Rather, we first multiply it
by a penalty
factor that depends on the number of points. This is in turn
determined as the 99\%
confidence level factor from a two sided Student-$t$ distribution.
Indeed, if the distribution of best-fit
$\as^{(k)}$ over replicas  is Gaussian, then the  difference between the sampled
and true central value follows a Student-$t$ distribution with
$N_\mathrm{rep}-1$ degrees of freedom, zero mean and scale parameter
$\Delta_{\sigma}/\sqrt{N_\mathrm{rep}}$ . A given confidence
level around the mean is  equal to  the standard deviation
$\Delta_{\sigma} T_{{\rm CL},(N_\mathrm{rep}-1)}$, where $ T_{{\rm CL},N}$
is  the percentile at CL confidence level for  the two-sided
confidence factor of the Student-$t$ distribution with
$N$ degrees of freedom.
Hence, we choose a $99\%$ confidence level, and we determine
$N_\mathrm{min }$  as the value yielding the minimum of
$\Delta_{\sigma} T_{0.99,(N_\mathrm{rep}-1)}$. Also in this case, the
dependence of results on the choice of selection criteria can be
studied a posteriori, and will be discussed in Sect.~\ref{sec:tests}.

\section{The strong coupling constant from NNPDF3.1}
\label{sec:results}

We now present the main result of this work, namely
the determination of $\amz$ based on the 
methodology discussed in Sect.~\ref{sec:fitsettings}.
We first present the best-fit result for $\alpha_s$ and its
experimental uncertainty, determined through  the correlated replica method.
We then discuss methodological and theoretical uncertainties.
We finally collect our final result and briefly compare it to other
recent determinations from PDF fits and to the PDG average.

\subsection{Best-fit results for $\alpha_s$ and statistical uncertainty}
\label{sec:finres}

We have determined $\amz$ both
at NLO and NNLO using the methodology and
dataset of the NNPDF3.1 global analysis~\cite{Ball:2017nwa}.
The only difference in the fit settings is the
theoretical description of the inclusive jet
production datasets at NNLO. Here we use 
exact NNLO theory~\cite{Currie:2016bfm} for the 
ATLAS~\cite{Aad:2014vwa} and CMS~\cite{Chatrchyan:2012bja}
inclusive jet measurements at 7 TeV, and discard the other jet datasets
used in NNPDF3.1 for which the NNLO calculation is not available (note
that, as in NNPDF3.1, only ATLAS data in the central rapidity bin are included).
To ensure a consistent comparison, the input datasets of the NLO and NNLO
fits are identical, up to small differences in the kinematical cuts
as explained in~\cite{Ball:2017nwa}. 

Specifically, we determine $\alpha_s$ by generating a set of 400 data 
replicas, and from them a set of 400 c-replicas each with 
21 values of $\alpha_s$, thus corresponding to a total of 8400
PDF replicas correlated as discussed in Sect.~\ref{sec:corrmc}.
These c-replicas are generated for $\amz$ ranging between 0.106 and
0.130, varied in
steps of $\Delta_{\as}=0.002$ between 0.106 and 0.112 and
between 0.128 and
0.130, and  in steps of $\Delta_{\as}=0.001$ between 0.112 and 0.128, adding
up to the total of 21 values. From these we determine $\as$ replicas, 
which form a representation of the probability distribution of $\as$.

At NNLO we find
\begin{equation}
\as^\text{NNLO}(m_Z) = 0.11845 \pm 0.00052~(0.4\%)\,.
\label{eq:finnnlo}
\end{equation}
This result
is  based on a total of $N_{\rm rep}=379$ c-replicas, selected from a
starting set of 400 after batch minimization
of three batches, using the minimization and selection methods described in
Sect.~\ref{sec:strategy}.
At NLO we find
\begin{equation}
\as^\text{NLO}(m_Z) = 0.12067 \pm 0.00064~(0.5\%) .
\label{eq:finnlo}
\end{equation}
In this case, the sample includes  $N_{\rm rep}=108$ c-replicas
selected after batch minimization with
two batches.
The smaller number of c-replicas selected at NLO is in part
explained by
the requirement (see Sect.~\ref{sec:strategy}) that  two
batches have converged for the given $\alpha_s$ value, which is of course
less severe when three batches are available, 
but the worse quality of the NLO fit also plays a role
since it causes more fits to be discarded by the post-selection criteria.

The uncertainty quoted in Eqs.~(\ref{eq:finnnlo}) and~(\ref{eq:finnlo}) is
that obtained using standard NNPDF methodology, namely, taking the standard
deviation over the $\alpha_s$ replica sample. We have verified that 
essentially the same results are obtained if instead we compute the 
68\% confidence interval.
The uncertainty is obtained in precisely the same way as our
PDF uncertainty, to which it is strongly correlated; it includes
the propagated correlated uncertainty from the underlying data, and
uncertainties coming from possible inefficiencies of the minimization
procedure.
This uncertainty is what we refer to as the experimental
uncertainty on $\amz$. It will have to be supplemented by 
methodological  and theoretical uncertainties, to be discussed
in Sects.~\ref{sec:tests} and~\ref{sec:method-th-uncertainties} below.

%%%%%%%%%%%%%%%%%%%%%%%%%%%%%%%%%%%%%%%%%%%%%%%%%%%%%%%%%%%%%%%%%%%%%
\begin{figure}[t]
\begin{center}
  \includegraphics[width=0.76\textwidth]{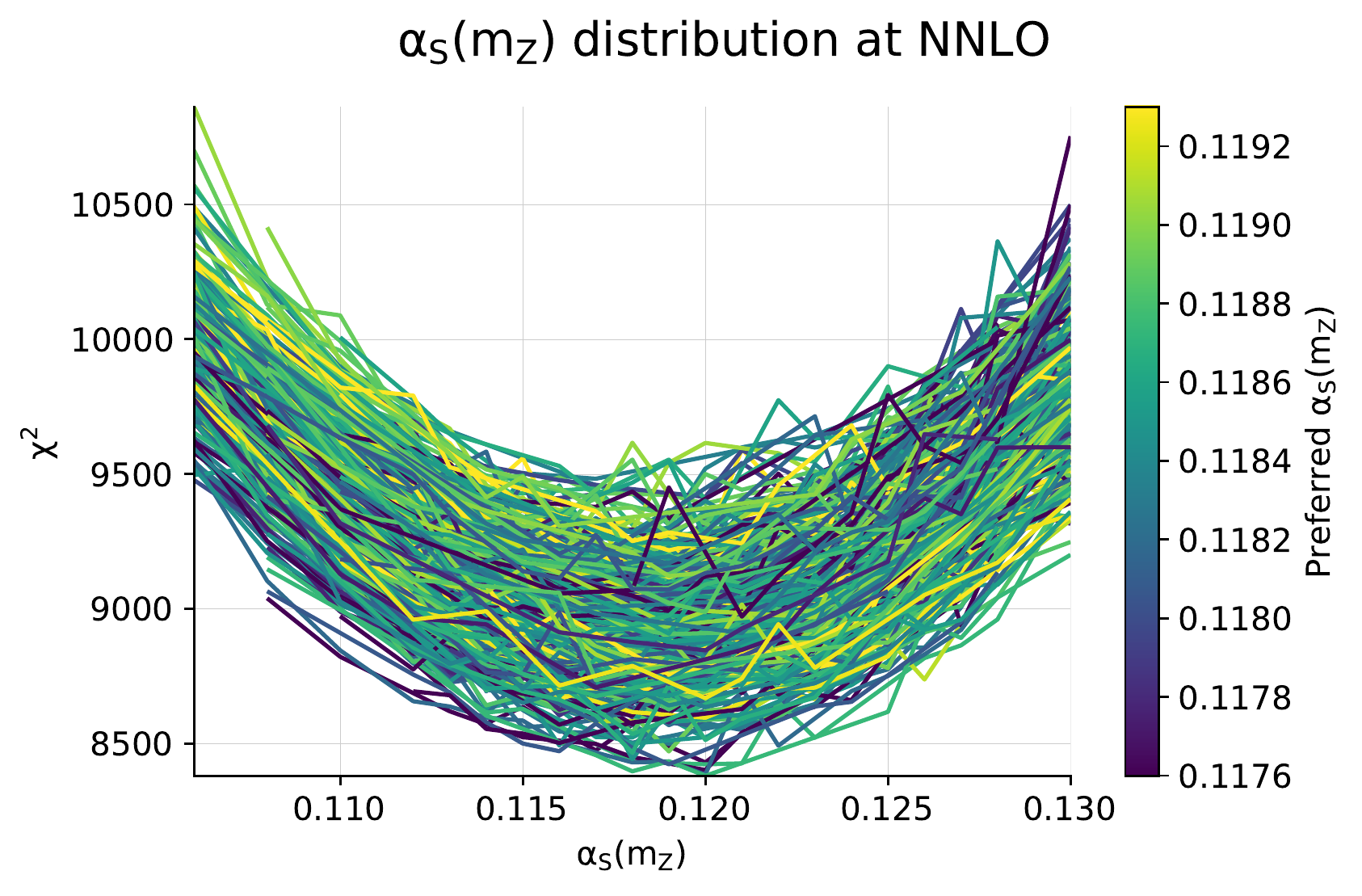}
  \caption{\small The $\chi^2$ profiles for each of the
$379$ c-replicas  used for
    the NNLO determination of $\alpha_s(m_Z)$, Eq.~(\ref{eq:finnnlo}).
    Each curve corresponds to an individual c-replica, and the 
 color scale indicates the best-fit $\as$ value
    determined from the parabolic fit to that curve.
   }
    \label{fig:nnlo_parabolas}
\end{center}
\end{figure}
%%%%%%%%%%%%%%%%%%%%%%%%%%%%%%%%%%%%%%%%%%%%%%%%%%%%%%%%%%%%%%%%%%%%%%

\begin{figure}[t]
\begin{center}
  \includegraphics[width=0.5\textwidth]{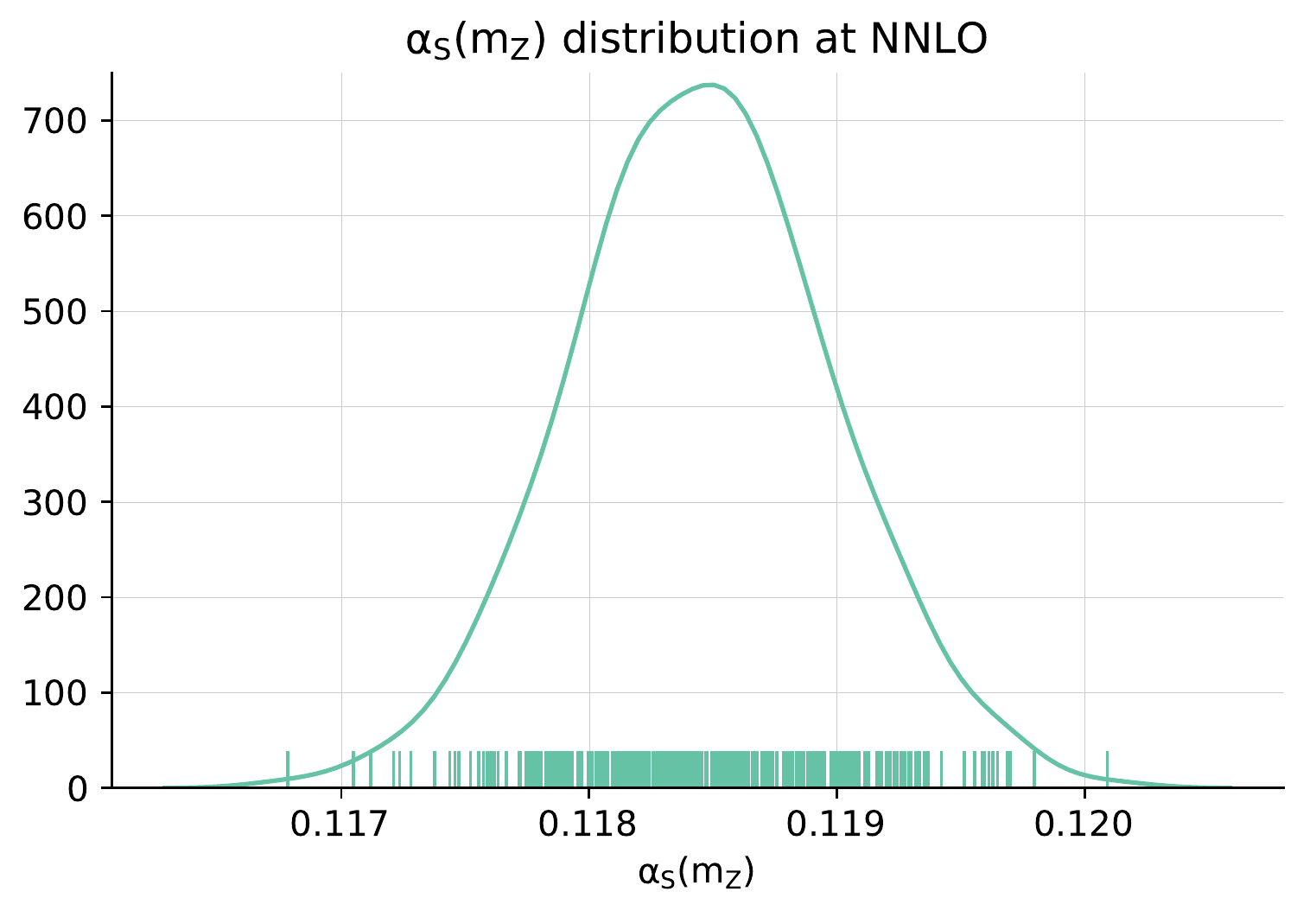}%
  \includegraphics[width=0.5\textwidth]{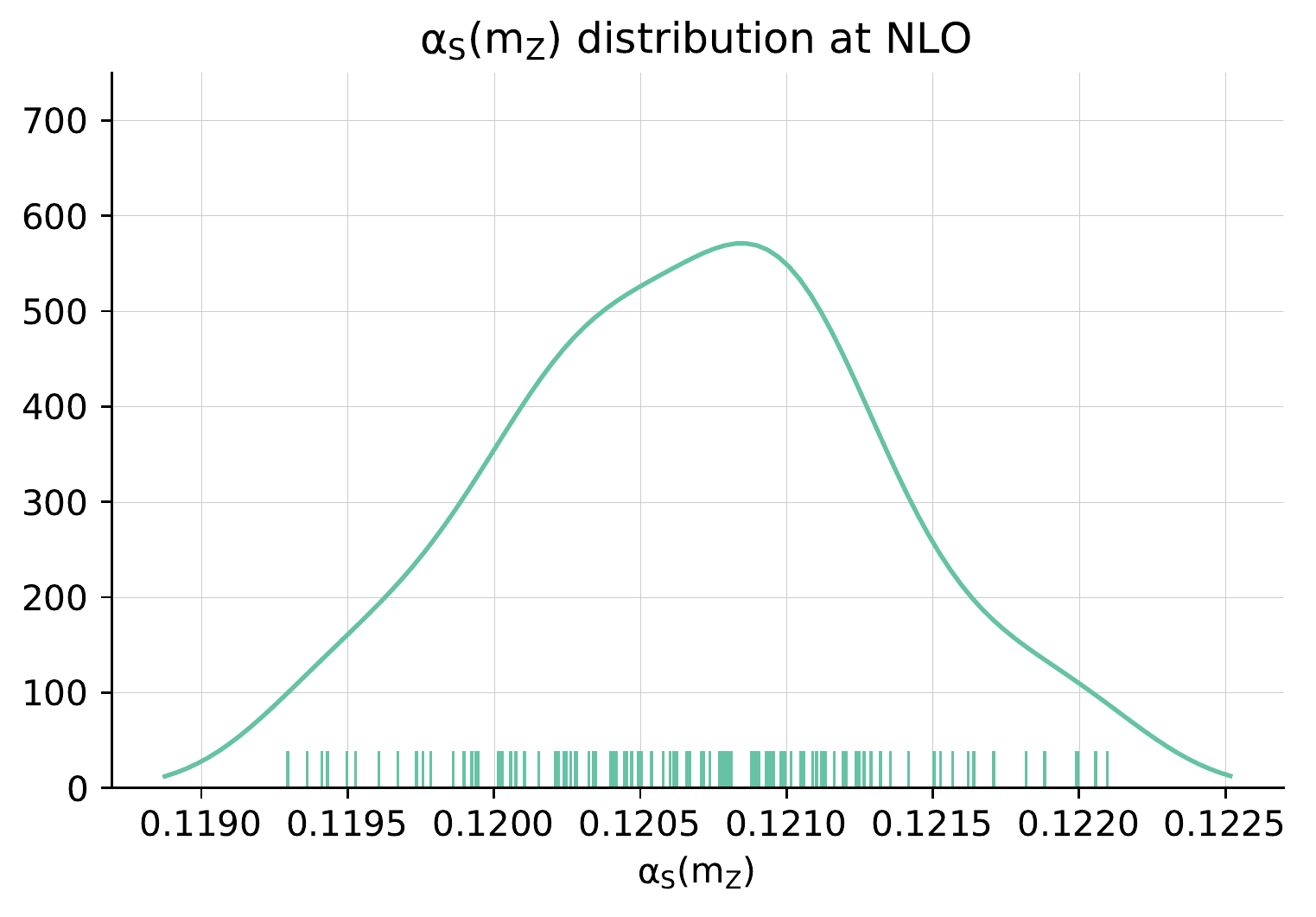}
  \caption{\small
The probability distributions for the best-fit $\alpha_s^{(k)}$ values
Eq.~(\ref{eq:alphakdef})
at NNLO (left) and at NLO (right). 
Each marker indicates the  $\alpha_s^{(k)}$ value corresponding to
 each individual c-replica.
    \label{fig:as_distributions}}
\end{center}
\end{figure}

The $379$ c-replicas selected  for
    the NNLO determination are shown in Fig~\ref{fig:nnlo_parabolas}.
    The color scale of each curve indicates the best-fit $\as$
    value. It is apparent
 that the vast  majority of the curves exhibit an approximately
    parabolic behaviour.
The probability distributions of the best-fit values $\alpha_s^{(k)}$ 
Eq.~(\ref{eq:alphakdef}) which correspond to each c-replica,
both at NLO and at NNLO, are shown in
Fig.~\ref{fig:as_distributions}, where the
markers indicate the value of $\alpha_s^{(k)}$ for each specific c-replica.
These probability densities have been determined using the
Kernel Density Estimate method, see~\cite{Carrazza:2016htc}.
We find that  the probability distribution
for $\amz$ is both shifted to higher values
and broadened when going from NNLO to NLO.
The decrease of the
best-fit value of  $\amz$ when going from NLO to NNLO
has been repeatedly observed before (see Table~1 of
Ref.~\cite{Martin:2009bu} for an extensive set of examples),
also in our previous determination~\cite{Lionetti:2011pw,Ball:2011us},
while the broadening is due to
the poorer quality of the NLO fit.

%%%%%%%%%%%%%%%%%%%%%%%%%%%%%%%%%%%%%%%%%%%%%%%%%%%%%%%%%%%%%%%%%%%%%
%%%%%%%%%%%%%%%%%%%%%%%%%%%%%%%%%%%%%%%%%%%%%%%%%%%%%%%%%%%%%%%%%%%%%%

The impact on the $\as$
determination of any subset of the input data can be  roughly assessed
by studying its contribution to the total figure of merit.
We have
done this by determining replica by
replica the corresponding partial $\chi^2_p$ for a process (or group
of processes) $p$, defined as 
the figure of merit Eq.~(\ref{eq:chi2not}) with the summation over
$i,j$ now restricted to data which belong to the specific subset $p$.
The $\alpha_s$ fit procedure through the correlated replica method is then
just repeated but using this partial $\chi^2_p$.
Namely, for each c-replica the partial  $\chi^{2(k)}_p$ for process $p$ is computed, a
parabola is fitted to it, the corresponding minimum $\alpha_{s,p}^{(k)}$  of the parabola
is determined, and  the resulting set of minima is used to find the value of
$\amz$ and its uncertainty.

\begin{table}[t]
  \begin{center}
    \renewcommand{\arraystretch}{1.2}
%%%%%%%%%%%%%%%%%%%%%%%%%%%%%%%%%%%%%%%%%%%%%%%%%%%%%%%%%%%%%
    \begin{tabular}{lrr}
\toprule
{} &   NLO &  NNLO \\
\midrule
Fixed-target charged lepton DIS       &   973 &   973 \\
Fixed-target neutrino DIS     &   908 &   908 \\
Collider DIS (HERA) &  1221 &  1211 \\
\midrule
Fixed Target Drell-Yan               &   189 &   189 \\
Collider Drell-Yan              &   378 &   388 \\
Inclusive jets                &   164 &   164 \\
$Z$ $p_T$                &   120 &   120 \\
Top quark pair production                &    26 &    26 \\
\bottomrule
Total               &  3979 &  3979 \\
\bottomrule
\end{tabular}
\vspace{0.3cm}
\caption{\small
\label{table:ndata}
Number of data points at 
NLO and NNLO corresponding to the different subsets of the input
experimental data considered here.
These eight subsets adds up to the total dataset.
}
\end{center}
\end{table}
%%%%%%%%%%%%%%%%%%%%%%%%%%%%%%%%%%%%%%%%%%%%%%%%%%%%%%%%%%%%%

Here we consider the following eight groups of processes $p$:
top production, the $Z$ $p_T$ distributions, collider and fixed target
Drell-Yan, inclusive jets, and deep-inelastic scattering (DIS)
either at HERA or at fixed-target experiments, in the latter case
separating charged lepton and neutrino beams.
The number of data points
corresponding to each of these data subsets is shown in
Table~\ref{table:ndata}.
 Not unexpectedly, the  $\chi^{2(k)}_p$
profiles for data subsets turn  out to be rather less parabolic than the total $\chi^2$, 
especially for processes such as neutrino
DIS or fixed target Drell-Yan that  have weak sensitivity to $\alpha_s$.

\begin{figure}[t]
\begin{center}
  \includegraphics[scale=0.54]{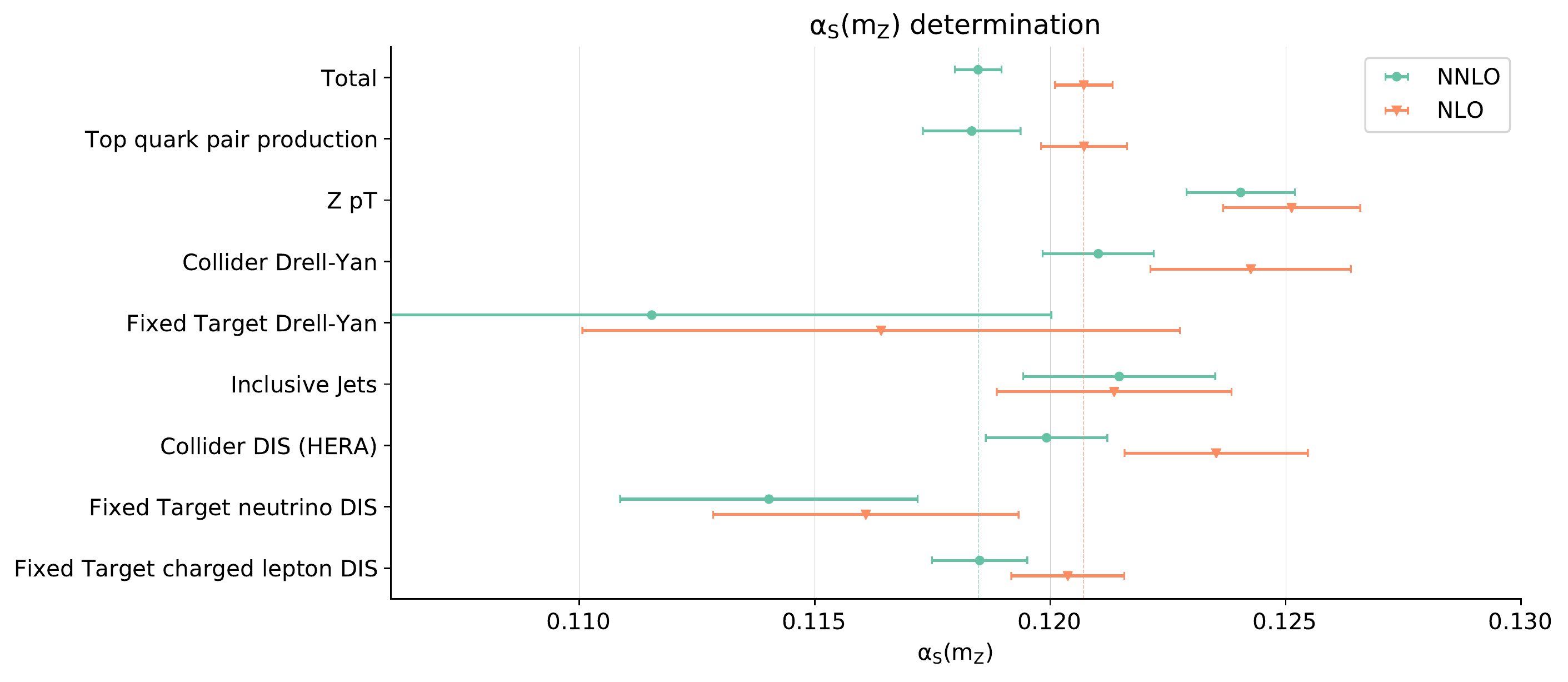}
  \caption{\small The  values of the partial $\amz$ and the
corresponding uncertainties determined from  $\chi^{2}_p$ for the
various families of processes $p$ of  Table~\ref{table:ndata}
 at NLO and NNLO.
    \label{fig:as_bestfit_replicas}
    }
\end{center}
\end{figure}

When determining $\amz$ from the partial $\chi^{2(k)}_p$,
we do not repeat the replica selection  and simply
use the same replicas selected for the total dataset.
Consequently, we must apply a form of post-selection,
whereby each time a parabola for $\chi^{2(k)}_p$ has no minimum the
corresponding c-replica is ignored.
At NNLO, for five out of eight data subsets we retain all 379 c-replicas, while
for jets, neutrino DIS, and fixed-target
Drell-Yan, we retain only 376, 366, and 302 c-replicas respectively.
The results for the partial $\amz$ determined from  $\chi^{2}_p$ for
the various families of processes are collected
in Fig.~\ref{fig:as_bestfit_replicas}.
The central value and
uncertainty shown are respectively determined as the median and 68\% symmetric
confidence level interval from the corresponding partial $\alpha_{s,p}^{(k)}$.
This is because
the analogue of Fig.~\ref{fig:as_distributions} for individual processes turns
out to be rather non-gaussian, especially for processes such as
fixed-target Drell-Yan that only have a weak handle on $\alpha_s$.

%%%%%%%%%%%%%%%%%%%%%%%%%%%%%%%%%%%%%%%%%%%%%%%%%%%%%%%%%%%%%%%%%%%%%         
%%%%%%%%%%%%%%%%%%%%%%%%%%%%%%%%%%%%%%%%%%%%%%%%%%%%%%%%%%%%%%%%%%%%%%        

The values of $\amz$ shown in  Fig.~\ref{fig:as_bestfit_replicas} should be
interpreted with some care.
Indeed,  the partial
$\chi^2_p$  is in each case computed using  PDF
c-replicas  determined from the minimization of the global  $\chi^2$.
These are in  general different from
the c-replicas that would be determined by simultaneous minimization of
$\chi^2_p$ with respect to $\alpha_s$ and the PDFs.
Therefore, the 
values of $\alpha_{s,p}$ in  Fig.~\ref{fig:as_bestfit_replicas} cannot be
interpreted as the best-fit values of $\amz$ for a given subset $p$.
They instead provide an estimate
of the pull on the best-fit $\amz$ value
that specific families of processes have within the global fit
subject to the constraints from the rest of the data.

Moreover, even their interpretation as pulls is only approximate.
Firstly, the replica selection is applied to the total $\chi^2$ rather than to
each partial $\chi^2_p$, so that
several partial $\chi^{2(k)}_p$ profiles turn out not to have a minimum.
Furthermore, the
total $\chi^2$ includes cross-correlations which are lost when
determining partial $\chi^2_p$, because the covariance matrix 
$C_{t_0}$ in Eq.~(\ref{eq:chi2not}) is generally nonzero
even when $i$ and $j$ belong to different data subsets.
For instance, inclusive jet,  $Z$ $p_T$, and Drell-Yan measurements
from the same experiment (ATLAS, or CMS) are correlated amongst themselves by the
common luminosity uncertainty.
Finally, partial $\alpha_s$ values are correlated through
the underlying PDFs, implying that the pulls should not be expected to
combine additively into the final result.

Even with all these caveats,  Fig.~\ref{fig:as_bestfit_replicas} 
shows that the very accurate $\amz$ value from the global
dataset is obtained from a combination of pulls
which correspond to values of $\amz$ dispersed about the global
best-fit value, without signs of tension or inconsistency, and subject to significant fluctuations
which are suppressed when constructing the total $\chi^2$.
This supports our conclusion that the current determination of $\amz$ from
a global fit is more precise and accurate than
determinations based on subsets of data relying on pre-existing PDF sets.

%%%%%%%%%%%%%%%%%%%%%%%%%%%%%%%%%%%%%%%%%%%%%%%%%%%%%%%%%%%%%%%%%%%%%
\begin{figure}[t]
\begin{center}
  \includegraphics[scale=0.47]{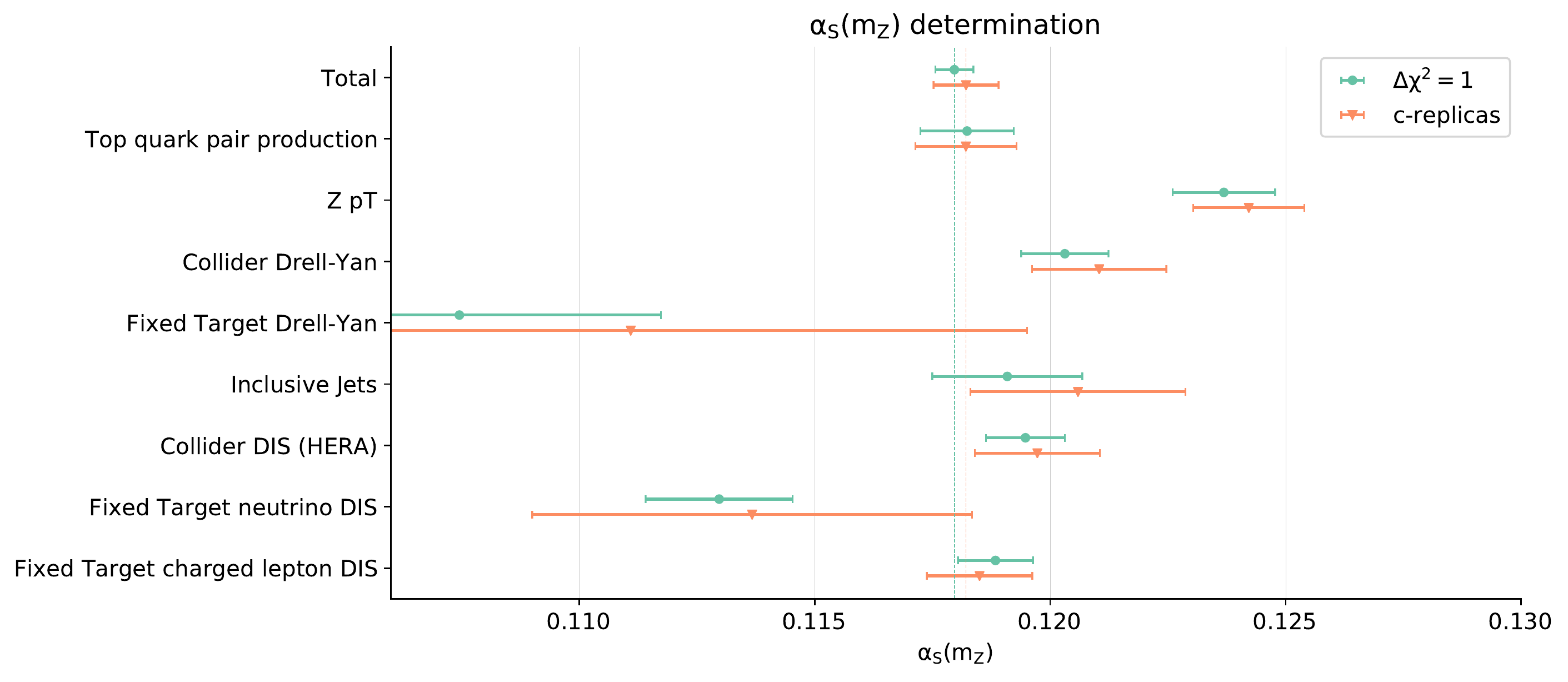}
  \caption{\small Comparison of the NNLO 
determination of $\amz$ using the method of~\cite{Lionetti:2011pw,Ball:2011us}, which neglects the
correlation between $\alpha_s$ and PDFs, and the current one based 
on the correlated replicas.
\label{fig:as_bestfit_NNLO_central}
}
\end{center}
\end{figure}
%%%%%%%%%%%%%%%%%%%%%%%%%%%%%%%%%%%%%%%%%%%%%%%%%%%%%%%%%%%%%%%%%%%%%%

Finally, we compare the current NNLO determination of $\amz$,
Eq.~(\ref{eq:finnnlo}) and Fig.~\ref{fig:as_bestfit_replicas},
with the one found using the
method of Refs.~\cite{Lionetti:2011pw,Ball:2011us}. We fix $\alpha_s$
and add  the contribution
to the $\chi^2$ from each standard PDF replica for that $\alpha_s$
value.  We then determine the total
$\chi^2(\alpha_s)$, 
fit a parabola to it, and 
determine the best-fit and uncertainty as the minimum and $\Delta
\chi^2=1$ interval.
For simplicity, we do this without using batch
minimization, i.e. we compute the total $\chi^2$ from one of the
batches (batch II, see Sect.~\ref{sec:tests} below) which then enter
the batch minimization procedure.
Using this method we find
\bea
\as^\text{NNLO}(m_Z) &=& 0.1180 \pm 0.0004~(0.3\%) \,,
\label{eq:globalnlo_central} \label{eq:globalnnlo_central} \\ \nonumber
\as^\text{NLO}(m_Z) &=& 0.1203 \pm 0.0004~(0.3\%)  \,.
\eea
Also in this case we can repeat the determination for different data
subsets based on the partial $\chi^2_p$, and
the corresponding results are compared in Fig.~\ref{fig:as_bestfit_NNLO_central}.

As expected, and discussed in the introduction and
in Sect.~\ref{sec:corrmc}, we find that the best-fit
values of $\amz$ determined with the old method~\cite{Lionetti:2011pw,Ball:2011us}
and with the new correlated
replica method are in good agreement, both for the global dataset and for
the data subsets.
The small differences in central values
are most likely due to uncertainties related to the
finite size of the replica sample, which, as
discussed in~\cite{Lionetti:2011pw,Ball:2011us}, can be
non-negligible when the old  method is used.
On the other hand, also as expected, neglecting the correlation between $\alpha_s$
and PDFs as in the old method leads in general to an underestimate of
the uncertainty on  $\alpha_s$.
This effect  is more marked
for processes such as  fixed-target Drell-Yan and neutrino DIS that
have  a limited  sensitivity to $\as$, because in this
case the difference in length of the semi-axes of the error ellipse in
Fig.~\ref{fig:ellipses} is large.

\begin{figure}[t]
\begin{center}
  \includegraphics[scale=0.45]{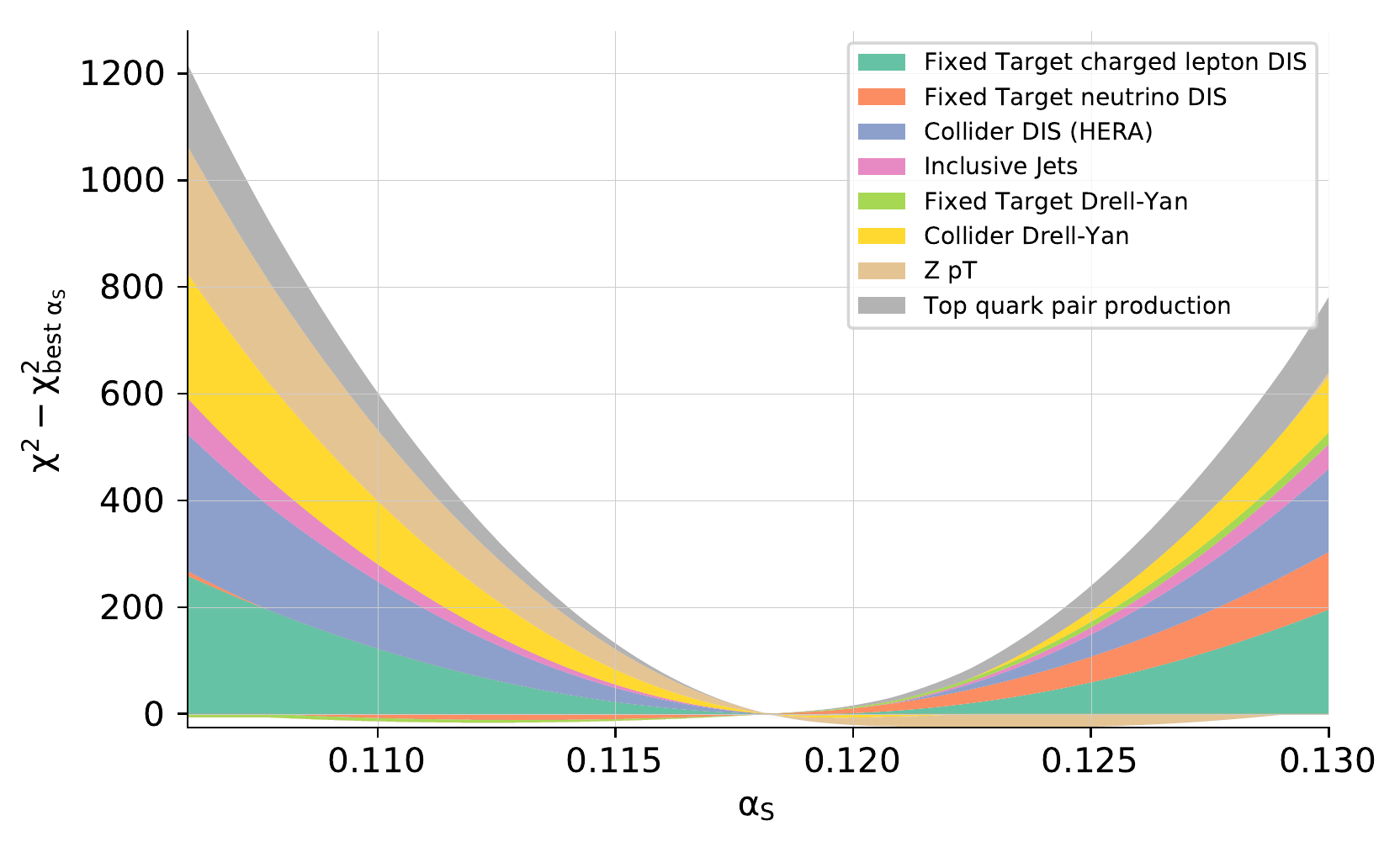}
  \includegraphics[scale=0.45]{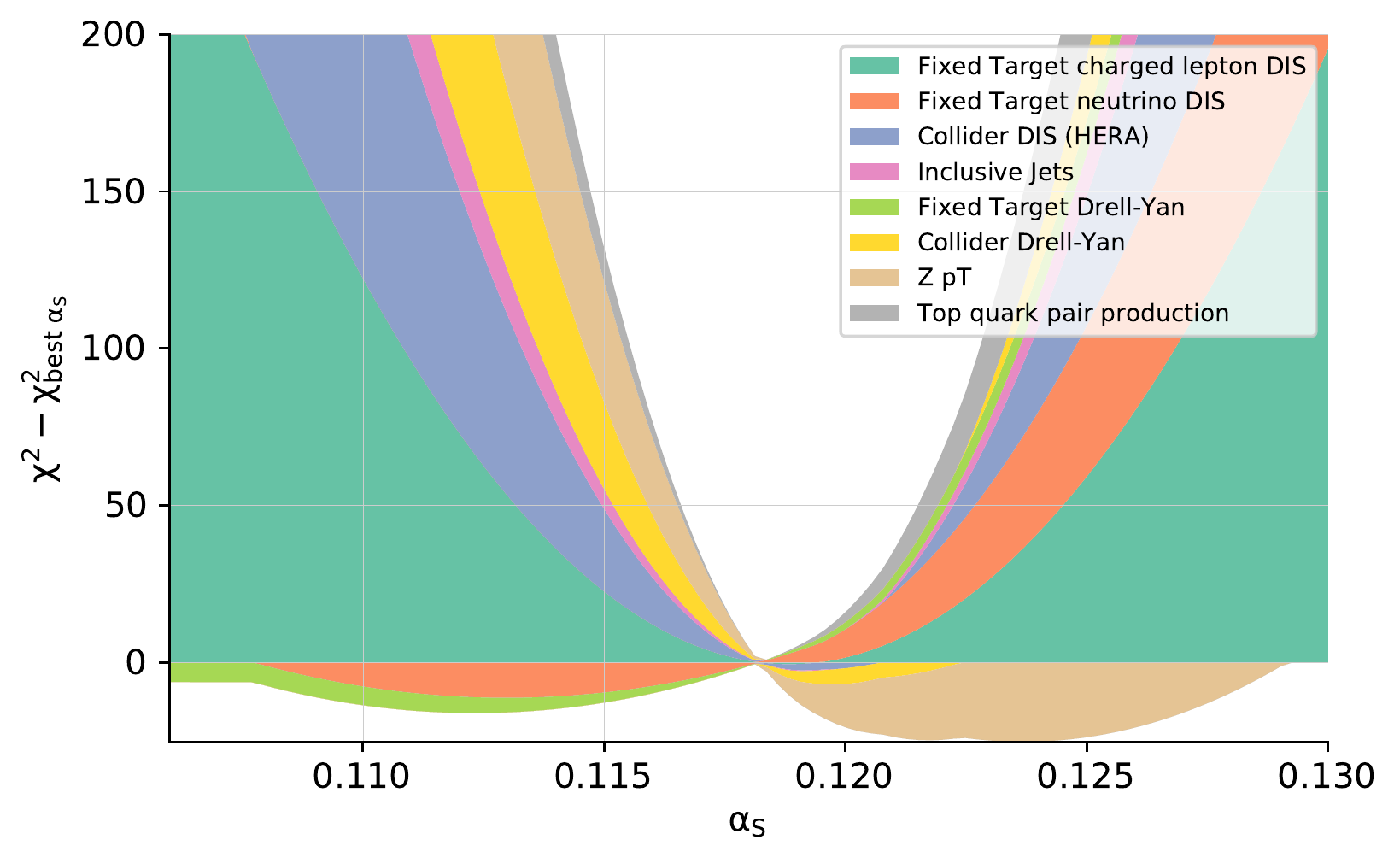}
  \caption{\small The NNLO cumulative differences,
$\chi^2_p(\as)-\chi^2_p(0.1185)$,
between the partial $\chi^2_p$ values
evaluated at $\amz$ and at best-fit value $\amz=0.1185$
    for different families of processes. In the part of the plot above
    (below) zero,
     only contributions from experiments for which the
    cumulative difference is positive (negative) are shown (see
    text). The plot is displayed either with a wider (left) or narrower
    (right) choice of range on the $y$ axis. 
}
    \label{fig:cumulative_chi2_diff}
\end{center}
\end{figure}

This determination of $\amz$  from the total $\chi^2$ also offers a
complementary way of quantifying how much each
family of processes constrains the final best-fit value, by
plotting the contribution of each data subset to the total $\chi^2$.
Specifically, we show in Fig.~\ref{fig:cumulative_chi2_diff} the  cumulative differences
at NNLO, 
$\chi^2_p(\as)-\chi^2_p(0.1185)$,
between each partial $\chi^2_p$ and its value computed at the global best-fit
$\amz$ value, neglecting cross-correlations between different data
subsets.
The plot is divided into two halfs: above zero,  only positive
differences are shown, and below zero, only negative
ones. Thus, when all differences are positive the plot shows the
breakdown of the total $\chi^2$ into the contribution of different
experiments (up to neglected cross-correlations), while when some of
them are negative the lower part of the plot shows by how much the
$\chi^2$ of the individual experiments shown has improved in
comparison to their 
value at the global minimum $\alpha_s(M_z)=0.1185)$. In order to
increase readability, the plot is displayed twice, with two different
choices of scale on the $y$ axis.

From this comparison, we observe that the LHC data
significantly contribute to constraining $\as$.
In particular, it is
interesting to note that the
$13$ data points from top-quark pair
production lead to a significant  contribution
to the total $\chi^2$ away from the best-fit, even though the global
dataset contains almost 4000 data points.
Similar considerations apply to the $Z$ $p_T$ distributions.
This means that  there is a small range of  values of $\as$ where these
two groups of processes are consistent with the rest of the data entering the fit,
thereby providing a tight constraint on $\as$.

%%%%%%%%%%%%%%%%%%%%%%%%%%%%%%%%%%%%%%%%%%%%%%%%%%%%%%%%%%%%%%%%%%%%%
%%%%%%%%%%%%%%%%%%%%%%%%%%%%%%%%%%%%%%%%%%%%%%%%%%%%%%%%%%%%%%%%%%%%%%

\subsection{Methodological uncertainties}
\label{sec:tests}

In view of the rather small  experimental
uncertainty on the final value of $\amz$,
Eqs.~(\ref{eq:finnnlo})--(\ref{eq:finnlo}), 
we need to assess possible uncertainties associated to the various
aspects of our methodology described in
Sect.~\ref{sec:fitsettings}.
Specifically, we discuss here the methodological uncertainties
associated to c-replica selection, batch minimization, the
quadratic approximation to $\chi^2$ profiles, and the treatment of
correlated systematics.

%%%%%%%%%%%%%%%%%%%%%%%%%%%%%%%%%%%%%%%%%%%%%%%%%%%%%%%%%%%%
\begin{table}[t]
  \begin{center}
     \renewcommand{\arraystretch}{1.29}
\begin{tabular}{c|c|c|c}
 $N_\text{min}$ & $\amz$ & $N_{\rm rep}$  & $\Delta_{\alpha_s}$    \\
\midrule

{18} & 					
$0.11842\pm 0.00031~(0.3\%)$  & 12  & 0.00009 \\
{15}    & $0.11844\pm
0.00044~(0.4\%)$     & 92  & 0.00005\\
{\bf 6}  &   $ {\bf 0.11845\pm
0.00052~(0.5\%)}$   & {\bf 379}  &  {\bf 0.00003} \\
{$3$}&  $0.11844 \pm 0.00056~(0.5\%)$  & 400 & 0.00003\\
\bottomrule
\end{tabular}
\vspace{0.3cm}
\caption{\small
  \label{table:discarded}
  Dependence of the NNLO determination of $\amz$ on the  minimum number of
  $\alpha_s$ values per c-replica $N_{\rm min}$ (see Sect.~\ref{sec:strategy}). 
 In each case, the best fit value
  and statistical uncertainty on $\alpha_s$ are shown, together with
  the  number of surviving c-replicas $N_{\rm rep}$ and the
  bootstrapping uncertainty $\Delta_{\alpha_s}$ Eq.~(\ref{eq:bootstrapmeansig}).
The value chosen using the selection criterion of  Sect.~\ref{sec:strategy},
which leads to the final vale of $\amz$ Eq.~(\ref{eq:finnnlo}), is $N_\text{min}=6$
(third row of the table, in boldface). 
}
\end{center}
\end{table}
%%%%%%%%%%%%%%%%%%%%%%%%%%%%%%%%%%%%%%%%%%%%%%%%%%%%%%%%%%%%

The replica selection algorithm determines an optimal value of
$N_\text{min}$, the minimal number of $\alpha_s$ for which results
must be available for a given c-replica to be selected.
We have varied this value from
its minimum $N_\text{min}=3$ (needed in order to fit
a parabola) to a high value  $N_\text{min}=18$ (meaning that at most
three values $\as$ can be missing in order for a c-replica to be retained).
Results for the number of c-replicas passing the criterion
and the ensuing value of $\alpha_s$ are collected in
Table~\ref{table:discarded} for a number of choices.
In each case we
also show the  finite-size
uncertainty  $\Delta_{\as}$ on the best-fit $\alpha_s$ estimated by
bootstrapping, Eq.~(\ref{eq:bootstrapmeansig}).

The number of surviving c-replicas varies significantly;
all the starting 400 c-replicas pass the loosest criterion
(i.e., it is
always possible to fit a parabola to any c-replica), but only
$N_\text{rep}=12$ c-replicas pass the most restrictive criterion.
However, even with this most restrictive criterion the finite-size
uncertainty is below the permille level.
For the value
selected by the algorithm, the finite-size uncertainty is of order
$0.03\%$, i.e. by almost a factor 20 smaller than the experimental
uncertainty Eq.~(\ref{eq:finnnlo}) and it does not decrease further even
when all c-replicas are kept.
The finite-size uncertainty on the $\alpha_s$ uncertainty  $\Delta_{\sigma}$ itself
Eq.~(\ref{eq:bootstrapstd}) is comparable in all cases. 

The value of $\amz$ and its experimental uncertainty are hence
very stable; the shift of central value and uncertainty when the
selection criterion is varied is always smaller than the finite-size
uncertainty.
This stability can be understood by observing that each
c-replica consists of at least $N_{\rm min}$ correlated  PDF replicas,
so each of the determinations shown in Table~\ref{table:discarded} is
obtained from more than $N_{\rm min}\times N_{\rm rep}$ PDF replicas.
 We thus estimate that the bootstrapping uncertainty, and
the related but smaller uncertainty due to choice of replica selection,
to be of order $\Delta_{\alpha_s}=0.00003~(0.03\%)$, one order of magnitude smaller
than the experimental uncertainty.

%%%%%%%%%%%%%%%%%%%%%%%%%%%%%%%%%%%%%%%%%%%%%%%%%%%%%%%%%%%%%%
\begin{table}[t]
  \begin{center}
    \renewcommand{\arraystretch}{1.29}
\begin{tabular}{l | c | c | c}
batches & $\amz$ & $N_{\rm min}$ & $N_{\rm rep}$ \\
\midrule
I           &  $0.11831\pm 0.00065~(0.5\%)$ & 9 & 310 \\
II           &  $0.11828\pm 0.00062~(0.5\%)$ & 14 & 216\\
III           &  $0.11822 \pm  0.00072~(0.6\%)$ & 13 & 369 \\
\midrule
I+II &   $0.11844 \pm 0.00054~(0.5\%)$ & 11 & 225 \\
I+III &  $0.11841 \pm  0.00058~(0.5\%)$ & 13 & 158 \\
II+II &  $0.11841 \pm  0.00060~(0.5\%)$ & 14 & 288\\
\midrule
{\bf I+II+III}     &    ${\bf 0.11845 \pm 0.00052~(0.4\%)}$ & {\bf 6} &
{\bf 379} \\
\bottomrule
\end{tabular}
\vspace{0.3cm}
\caption{\small
  \label{table:batchmin}
  Results for the NNLO determinations of $\amz$
  using different combinations of the three available batches.
  In each
  case we show both the best-fit value of $\amz$, the  minimum number of
  $\alpha_s$ values per c-replica $N_{\rm min}$,
  and the corresponding number
  surviving c-replicas $N_{\rm rep}$.
  The last row (in boldface) corresponds to our final result Eq.~(\ref{eq:finnnlo}).
  }
\end{center}
\end{table}
%%%%%%%%%%%%%%%%%%%%%%%%%%%%%%%%%%%%%%%%%%%%%%%%%%%%%%%%%%%%%%

We next turn to discuss batch minimization.
The results shown in
Table~\ref{table:discarded} all correspond  to the NNLO baseline
which uses batch minimization with three batches.
In order to assess
the impact of batch minimization,
in Table~\ref{table:batchmin} we compare results obtained with each of
the three batches, with the three possible pairs, and combining the
three batches.
In each case we show the final best-fit $\amz$ and
experimental uncertainty, the value of $N_{\rm min}$, the
minimum number of $\alpha_s^{(k)}$  values per c-replica, and the number
of surviving c-replicas $N_{\rm rep}$.

It is clear from this comparison that as more batches
are combined, results become more stable. The
values of $N_{\rm min}$ are on average larger with two batches, and
 larger still with three, but without a reduction of the number of
surviving c-replicas $N_{\rm rep}$ as was observed in
Table~\ref{table:discarded}. With three batches,  $N_{\rm rep}$
is largest even though  $N_{\rm min}$ is also largest.
This means that,
thanks to batch minimization, the number of available  $\alpha_s^{(k)}$ values
per replica is on average higher.
It follows that the finite-size
uncertainty is reduced by batch minimization, thus leading to the
very small uncertainties shown in Table~\ref{table:discarded}.

The values of $\amz$ behave as expected upon use of batch
minimization. The experimental uncertainty is reduced when more
batches are used and the central values with different combinations of
batches are all consistent with each other within given uncertainties. Furthermore,
the differences in central values with different combinations of
batches are reduced upon use of batch minimization (they are smaller
when using two batches than when using a single batch). Additionally, the shift in
central value when increasing the number of batches is rather smaller
than the uncertainty, and, finally, the central value is stabilized when
increasing the number of batches, so the difference between two and
three batches is on average smaller than the difference between one
and two batches.

%%%%%%%%%%%%%%%%%%%%%%%%%%%%%%%%%%%%%%%%%%%%%
\begin{table}[t]
  \begin{center}
         \renewcommand{\arraystretch}{1.29}
\begin{tabular}{l|c|c|c}
$N_{\rm trim}$ & fitted $\amz$ range & $\amz$  & $N_{\rm rep}$  \\
\midrule
      {\bf 0} &  ${\bf [0.106,0.130]}$  & $ {\bf 0.11845\pm
        0.00052~(0.4\%)}$  & {\bf 379} \\
2  &  $[0.108,0.128]$  &$0.11846 \pm 0.00045~(0.4\%)$ & 218 \\
5 &  $[0.110,0.126]$  & $0.11852\pm  0.00051~(0.4\%)$ & 290 \\
10 &  $[0.114,0.124]$  & $0.11869 \pm 0.00046~(0.4\%)$ & 32 \\
15 &  $[0.115,0.120]$  &$0.11822 \pm 0.00079~(0.7\%)$ & 10 \\
\midrule
4 &  $[0.113,0.130]$  &$0.11850 \pm 0.00058~(0.5\%)$ & 296 \\
5 &  $[0.106,0.124]$  &$0.11855 \pm 0.00059~(0.5\%)$ & 197 \\
\bottomrule
\end{tabular}
\vspace{0.3cm}
\caption{\small
  \label{table:farvariation} Results for the NNLO determinations of $\amz$
  when the $N_{\rm trim}$ outer values of  $\as$  are not used and the
  fit is restricted to a smaller range.
  In the bottom part of the
  table we also show results found discarding values asymmetrically,
  at the upper or lower edge of the range.
  In each
  case we show the number of discarded $\as$ values,
the best-fit value of $\amz$, and the number of
  surviving c-replicas $N_{\rm rep}$.
  The first row (in boldface) corresponds to our final result Eq.~(\ref{eq:finnnlo}).
}
\end{center}
\end{table}
%%%%%%%%%%%%%%%%%%%%%%%%%%%%%%%%%%%%%%%%%%%%%

We conclude that the value of $\amz$ found using three  
batches is the most accurate.
We observe that even the shift
between the three-batch value and the single-batch value which differs most from
it is about a  third of the finite-size uncertainty. We take this as
further evidence that there is no extra contribution of methodological
origin due to batch minimization to be added to the statistical
uncertainty. We finally observe that the two-batch result is in fact
consistent within its very slightly larger uncertainty, thus
justifying the use of only two batches at NLO.

We next turn to the methodological uncertainties related to the quadratic fitting of
$\chi^2$ profiles.
We have studied this in three different ways: by
removing outer values of $\amz$ from the fit; by adding higher order
terms to the fitting function; and by changing the fitting
variable. We discuss each in turn.

First, we have repeated the NNLO
determination removing  $\as$ values that are farthest
from the best-fit value $\amz=0.1185$, fitting a smaller range of
values around the minimum. 
As a further consistency check, we have removed $\as$ values
asymmetrically.
Results are shown 
in Table~\ref{table:farvariation}; in each case we show the
number  of discarded 
$\alpha_s$ values $N_{\rm trim}$,
the resulting fitted range, the best fit $\amz$ and
uncertainty, and the number of surviving c-replicas $N_{\rm rep}$.
Here too, the behaviour is consistent with
expectations. As the fitted range is reduced, the experimental
uncertainty increases and the number of surviving c-replicas decreases
(thereby also increasing the finite-size uncertainty).
The
central value, however, is extremely stable; the shift in central value when
restricting the range is always more than a factor two smaller than
the experimental uncertainty.
In fact, 
the shift is never larger than
$\Delta=0.00010~(0.08\%)$ unless the  number of surviving c-replicas
becomes of order ten, in which case the finite-size uncertainty
(recall Table~\ref{table:discarded}) is of the same order or
larger.

A different way of testing for deviations from quadratic behaviour is
to apply a criterion to assess fit quality to both quadratic and
cubic fits.
Here we use the Akaike Information Criterion
(AIC)\cite{akaike1974new}, which estimates the
expected relative distance between a given fitted model and the unknown
underlying law~\cite{burnham2004multimodel}.
The AIC score balances
goodness of fit against simplicity of the model. A  lower score
corresponds to a lower expected distance measured by the
Kullback–-Leibler divergence.
The AIC score is 
defined by
\begin{equation}
	\label{eq:aic}
	\mathrm{AIC} = 2r - 2\ln{L} + \frac{2r(r+1)}{n-r-1} \, ,
\end{equation}
where $r$ is the number of degrees of freedom of the model, $n$ is the number of
fitted points, and $\ln(L)$ is the log-likelihood associated with the
model.

In our case, we fit to $\chi^{2(k)}(\alpha_s)$,
Eq.~(\ref{eq:chi2kdef}),  viewed as a function of $\alpha_s$ using either a
parabola (as in 
our default determination) or a higher order polynomial.
The
log-likelihood is then in each case just the $\chi^2$ of this fit.
Computing  the AIC score for each fitted profile, 
averaging over c-replicas, and taking the variance of results as a
measure of the uncertainty, we find 
$\mathrm{AIC}=169 \pm 37$ for the default quadratic fit
and  $\mathrm{AIC}=173 \pm 35$ for a cubic fit.
We conclude that there is no evidence that a cubic fit is better than
a quadratic one.

%%%%%%%%%%%%%%%%%%%%%%%%%%%%%%%%%%%%%%%%%%%%%%%%%%%%%%%%%%%
\begin{table}[t]
  \begin{center}
          \renewcommand{\arraystretch}{1.29}
\begin{tabular}{l|c|c}
  & $\amz$  & $N_{\rm rep}$  \\
\toprule
{\bf default} & $ {\bf 0.11845 \pm 0.00052~(0.4\%)}$ & {\bf 379} \\
ln & $0.11845 \pm 0.00052~(0.4\%)$ & 379 \\
exp     & $0.11849 \pm 0.00052~(0.4\%)$ & 379 \\
\bottomrule
\end{tabular}
\vspace{0.3cm}
\caption{\small
  \label{table:explog}
  Same as Table~\ref{table:discarded}, comparing
  the default parabolic fitting (in boldface) of the $\chi^2(\as)$ profiles
  with those with a transformed input, both
  $\chi^2\lp \ln(1+\as)\rp $ and $\chi^2\lp \exp(\as)\rp$.
}
\end{center}
\end{table}
%%%%%%%%%%%%%%%%%%%%%%%%%%%%%%%%%%%%%%%%%%%%%%%%%%%%%%%%%%%

We perform a final test based on the observation that any transformation of
the error function profile of the form
\be
\label{eq:transformedchi2profiles}
\chi^{2}(\as)\to\chi^{2}(f(\as)) \, ,
\ee
where $f$ is sufficiently smooth and monotonic,
should lead to the same best-fit value of $\as$.
The results of fitting $\as$
from the transformed profiles
Eq.~(\ref{eq:transformedchi2profiles})
with
$f(\as)=\exp(\as)$ and $f(\as)=\ln(1+\as)$ 
are shown in Table~\ref{table:explog}.
The argument of
 the log is shifted so that $f(\as)$ admits a Taylor expansion in
 powers of $\as$.
 Reassuringly, we find extreme stability with respect to these transformations
 of the fitting argument.

Combining results from
Tables~\ref{table:farvariation} and~\ref{table:explog} and the analysis based
on the AIC score we can conservatively take as an estimate of the uncertainty related to parabolic
fitting the largest shift observed in Table~\ref{table:discarded},
neglecting the cases with $N_{\rm rep}<100$ which are dominated by
finite-size uncertainty, namely
\be
\Delta_{\rm par}=0.00010~(0.08\%).
\label{eq:parunc}
\ee

We finally turn to the uncertainty related to the treatment of
experimental correlated systematic errors.
As mentioned in Sec.~\ref{sec:corrmc},
the covariance matrix in the presence of multiplicative uncertainties
should be identified with the experimental covariance matrix, in order
to avoid biasing the fit~\cite{D'Agostini:1993uj}.
We thus adopt the $t_0$ method, introduced
in~\cite{Ball:2009qv}, benchmarked in~\cite{Ball:2012wy}, and
used for the determination of all NNPDF sets from NNPDF2.0~\cite{Ball:2010de} onwards.
In this procedure, the normalization of the multiplicative uncertainties
that enter the  covariance matrix is iteratively determined from a  prior theory prediction.
Because the PDFs and $\alpha_s$
are now determined on  the same footing, the same covariance matrix
is used for both. It is clear that the
same $\chi^2$ definition must be used in Eq.~(\ref{eq:prob}) as in
Eqs.~(\ref{eq:jointprobf})-(\ref{eq:jointproba}) in order for the same minimum
to be found.

\begin{figure}[t]
\begin{center}
  \includegraphics[scale=0.5]{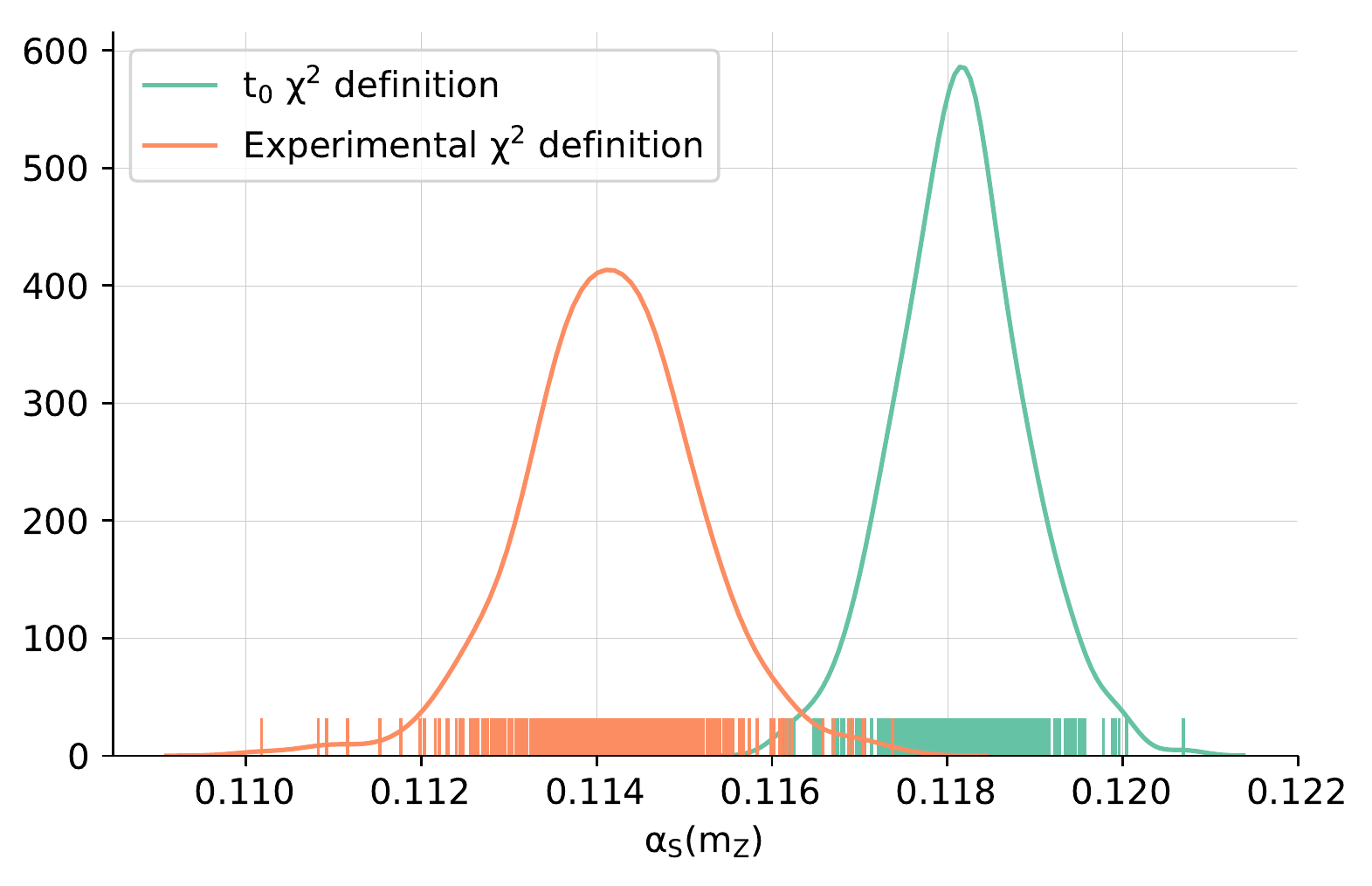}\\
  \includegraphics[scale=0.47]{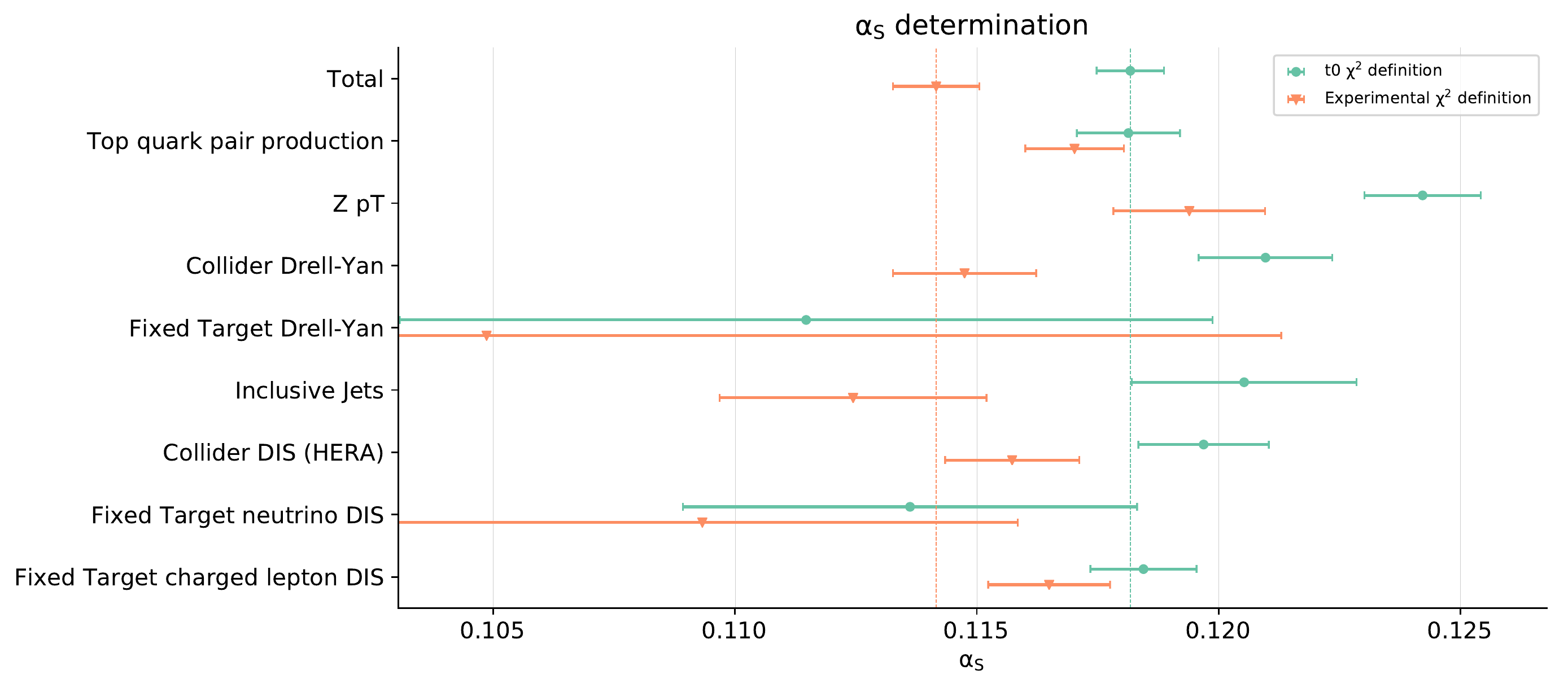}
  \caption{\small Top: probability distributions for the best-fit
    $\alpha_s^{(k)}$ values (same
    as Fig.~\ref{fig:as_distributions})  and bottom: values of the
    partial $\amz$  and  corresponding uncertainties (same as  
Fig.~\ref{fig:as_bestfit_NNLO_central}) in both cases comparing  NNLO
results from a single batch found  using either a
    consistent or an inconsistent definition of the $\chi^2$.
  }
	\label{fig:expchi2dist}
\end{center}
\end{figure}

Indeed, it is interesting to note that using an inconsistent definition of
the covariance matrix significantly biases the result of $\amz$.
In Fig.~\ref{fig:expchi2dist} we compare the distribution of 
NNLO $\amz$
values  as well as the total and partial best-fit values and
uncertainties,
 computed for a single batch,  either consistently using
the $t_0$ covariance matrix (see
Figs.~\ref{fig:as_distributions},~\ref{fig:as_bestfit_NNLO_central} for
the corresponding results with three batches) or inconsistently using
the experimental 
covariance matrix. 
We find that the inconsistent definition leads to a much broader
distribution for the total $\chi^2$, thereby signaling the lack of
consistency, and, more 
importantly, a biased central value
$\as(m_Z) = 0.114 \pm 0.001^\text{exp}~(0.9\%)$, shifted 
by  about 9-$\sigma$  in comparison to the correct result
Eq.~(\ref{eq:finnnlo}).
The fact that a downward shift of $\amz$ is observed when using the
inconsistent definition can be understood based on the observation that
the bias~\cite{D'Agostini:1993uj} typically leads to the best-fit
undershooting the data, essentially because with multiplicative uncertainties a
lower prediction has a smaller uncertainty~\cite{dagos}.  Indeed,
inspection of the partial best-fit values shows that the bias is much
stronger for  
collider experiments than the fixed-target ones. 
This is what one would expect, because systematic uncertainties are
multiplicative for collider experiments, while they are mostly
additive for fixed-target~\cite{Ball:2012wy}, so any effect or bias
related to the 
treatment of multiplicative uncertainties should be mostly seen in
collider data.

%%%%%%%%%%%%%%%%%%%%%%%%%%%%%%%%%%%%%%%%%%%%%%%%%%%%%%%%%%%%%%%%%%%%%
%%%%%%%%%%%%%%%%%%%%%%%%%%%%%%%%%%%%%%%%%%%%%%%%%%%%%%%%%%%%%%%%%%%%%%

%%%%%%%%%%%%%%%%%%%%%%%%%%%%%%%%%%%%%%%%%%%%%%%%%%%%%%%%%%%%%%%%
\begin{table}[t]
  \begin{center}
     \renewcommand{\arraystretch}{1.29}
\begin{tabular}{l|c|c}
\toprule
$t_0$      & $\amz$                        & $N_{\rm rep}$ \\
\midrule
 I     & $0.11844 \pm 0.00052 (0.4\%)$ & 379 \\
 II    & $0.11845 \pm 0.00052 (0.4\%)$ & 379 \\
 III   & $0.11841 \pm 0.00051 (0.4\%)$ & 356 \\
\bottomrule
\end{tabular}
 \vspace{0.34cm}
\caption{\small
\label{table:t0variations}
Best-fit value of $\amz$ and experimental uncertainty found using
three different forms of the $t_0$ covariance matrix (see text); the
second row corresponds to the central result Eq.~(\ref{eq:finnnlo}). The
number of  c-replicas selected in each case is also shown.  
}
\end{center}
\end{table}
%%%%%%%%%%%%%%%%%%%%%%%%%%%%%%%%%%%%%%%%%%%%%%%%%%%%%%%%%%%%%%%%%%

The use of the $t_0$ procedure in principle leads to a further
methodological uncertainty related to the choice of the prior used for
the construction of the $t_0$ matrix, which should therefore be assessed.
In order to determine the final result
Eq.~(\ref{eq:finnnlo}) the $t_0$ matrix was constructed using the
best-fit PDF set from batch~II of Table~\ref{table:batchmin}. We have
repeated the determination constructing the $t_0$ matrix from the
best-fit  PDF set of either of the other two batches.
Results are
collected in  Table~\ref{table:t0variations}. It is clear that, using
the consistent $t_0$ method, results are extremely stable.
We can conservatively estimate the uncertainty due to the choice of
$t_0$ from the largest shift seen in Table~\ref{table:t0variations} as
$\Delta_{t_0}=0.00004~(0.03\%)$.

In summary, we conservatively estimate methodological uncertainties by
adding in quadrature  the finite-size uncertainty
$\Delta_{\alpha_s}=0.00003$, the uncertainty related to the
parabolic approximation  $\Delta_{\rm par}=0.00010$ and the
uncertainty related to the treatment of correlated systematics
$\Delta_{t_0}=0.00004$, with the result
\be
\sigma^\text{meth} = 0.00011~(0.09\%) \, .
\label{eq:NNLO_meth}
\ee
Therefore, we find  that, at NNLO, methodological uncertainties 
are smaller than the experimental uncertainties Eq.~(\ref{eq:finnnlo}) 
by a factor five.

\subsection{Theoretical uncertainties from missing higher orders}
\label{sec:method-th-uncertainties}

A determination of $\amz$ is  dependent on the
perturbative order of the QCD calculations on which it
relies. Therefore, at any fixed order it is affected by a missing
higher order uncertainty (MHOU).
In older, and also some more recent  determinations of $\amz$
(specifically for determination in PDF fits see
Refs.~\cite{Lionetti:2011pw,Harland-Lang:2015nxa,Alekhin:2017kpj}) 
 no attempt was made to estimate the MHOU, 
and sometimes NLO or NNLO values of $\amz$
 were quoted with the
understanding that they might differ by an amount greater than the
quoted uncertainty due to this missing uncertainty.
However, as the experimental
uncertainty decreases, an estimate of the MHOU becomes mandatory, and in
the context of PDF fits it was done e.g. in Ref.~~\cite{Ball:2011us}.
 Indeed, this uncertainty, usually estimated by
scale variation, is
typically dominant in more recent determinations~\cite{Johnson:2017ttl,Andreev:2017vxu,Klijnsma:2017eqp,Aaboud:2017fml,
  Bouzid:2017uak,Chatrchyan:2013haa,Britzger:2017maj}.

In the present case, a first handle on the MHOU associated
to $\alpha_s$ is provided
by the difference between the NLO and NNLO results 
Eqs.~(\ref{eq:finnnlo}) and~(\ref{eq:finnlo}), namely
\be
\Delta\as^\text{pert} \equiv | \as^\text{NNLO} - \as^\text{NLO} | = 0.0022 \, ,
\label{eq:MHOU}
\ee
which corresponds to a $2\%$ shift of the NNLO central value.
This is about four times
larger than the experimental uncertainty in
Eq.~(\ref{eq:finnnlo}), thereby  suggesting that even at  NNLO
the  MHOU on the  $\amz$ determination might be comparable to, or larger
than the experimental uncertainty.

In our previous  determination of $\alpha_s$ Ref.~\cite{Ball:2011us}
the MHOU was estimated using the Cacciari-Houdeau (CH)
method~\cite{Cacciari:2011ze}, which relies on a
Bayesian estimate of the missing higher perturbative orders based on
the behaviour of the 
known orders. Use of exactly the same method of
Ref.~\cite{Ball:2011us}, to which the reader is referred for details,
leads to the values 
\begin{align}
\Delta^\text{CH,\,NLO}&= 0.003 \label{eq:MHOUn} \, , \\ 
\Delta^\text{CH,\,NNLO}&=0.0004 \,  \label{eq:MHOUnn}
\end{align}
 for the 68\% confidence level MHOU on $\alpha_s(M_Z)$.
The rather large difference in the MHOU estimate between NLO and NNLO
stems from the fact that 
there is a significant shift when going from LO to NLO, but a much smaller one when
going from NLO to NNLO.

The NLO estimate of the MHOUs in Eq.~(\ref{eq:MHOUn}) is reassuringly in good
agreement with the observed shift Eq.~(\ref{eq:MHOU}).
The NNLO uncertainty Eq.~(\ref{eq:MHOUnn}) is also consistent with
expectations based on the CH uncertainty estimate of 
Ref.~\cite{Ball:2011us}, where the value of $\amz$ determined using
the NNPDF2.1 set was found to lead to 
$\Delta^\text{CH,\,NNLO}=0.0009$. Indeed, PDF uncertainties in the
NNPDF3.1 set are generally smaller than those on NNPDF2.1 by a
 factor of two or more, due to significant impact  of LHC data in the
 more recent determination.
 In addition,  the shift between NLO and NNLO PDFs
is found to be smaller in NNPDF3.1 than in previous NNPDF
sets~\cite{fortetalk}, presumably 
because MHO terms pull in different directions and thus partly cancel
each other to a greater extent in a more global fit.
Indeed, we find a
similar increase  of perturbative
stability  of PDFs and of the associated $\amz$
by repeating the analysis presented here for reduced datasets~\cite{nnpdflh}.
Therefore, the reduction of the MHOU by a
comparable factor in  Eq.~(\ref{eq:MHOU}) in comparison to
Ref.~\cite{Ball:2011us} is expected.

Nevertheless, 
the very small value of the MHOU at NNLO, Eq.~(\ref{eq:MHOUnn}),
even smaller than the already small experimental uncertainty
Eq.~(\ref{eq:finnnlo}), may seem rather too optimistic. There are
furthermore several reasons of principle and practice why the
reliability of the CH method in the present case is dubious. The main
one is that the implementation of the  method suggested in
Ref.\cite{Ball:2011us} relies on a guess for an underlying ``true''
value  $\as^{(0)}$, and for a leading-order value 
$\as^\text{LO}$, neither of which
is known. The result Eqs.~(\ref{eq:MHOUn}-\ref{eq:MHOUnn}) is 
obtained by varying $\as^\text{LO}\in[0.10,0.14]$.
 and $\as^{(0)}\in [0.110,0.125]$. These are however largely arbitrary
 choices, and the final answer relies on them.

We therefore prefer to adopt a more conservative estimate.
Namely, we assume that the MHOU on the NNLO result is 
half the difference  between the NLO and NNLO results  Eq.~(\ref{eq:MHOU}):
\be
\Delta\alpha^{\rm th}_s =0.0011~(0.9\%)\, ,
\label{eq:NNLO_MHOU}
\ee
about  twice the size of the corresponding experimental
uncertainty  Eq.~(\ref{eq:finnnlo}). Whereas this is surely a very
crude estimate, we do not feel that any of the available methods can
lead to a more reliable conclusion.

On top of the missing higher fixed-order QCD corrections, 
several other aspects of the theory used in the simultaneous determination of $\amz$ and PDFs
also lead to  uncertainties. These include the values of the heavy quark masses,
standard model parameters (specifically CKM matrix elements and
electroweak couplings), electroweak corrections, QCD resummation
corrections~\cite{Bonvini:2015ira,Ball:2017otu}, QCD power corrections,
and nuclear corrections.
Many of these uncertainties were 
assessed in the NNPDF3.1 PDF determination that we are relying
upon~\cite{Ball:2017nwa}, and found to be smaller than PDF
uncertainties. In particular, the dependence on the charm mass in
previous PDF determinations is substantially reduced 
in NNPDF3.1 and likely rather smaller than the MHOU, thanks to the
presence of an independently parametrized charm PDF~\cite{Ball:2016neh},
and electroweak
corrections are carefully kept under control thanks to the choice of
suitable kinematic cuts.
But PDF uncertainties  mix with the experimental
uncertainty on $\amz$, with which they are strongly correlated, and
are in fact indistinguishable from it, as discussed in
Sect.~\ref{sec:corrmc}, so the hierarchy of uncertainties on PDFs and
$\amz$ is the same. We conclude that we have evidence that most of
these theoretical uncertainties are
sub-dominant in comparison to the experimental uncertainty
Eq.~(\ref{eq:finnnlo}), and thus even more so in comparison to the
MHOU Eq.~(\ref{eq:NNLO_MHOU}).

\subsection{Final results and comparisons}
\label{subsec:final}

We can now collect results. Combining the 
NNLO value and  experimental uncertainty Eq.~(\ref{eq:finnnlo}),
the methodological uncertainty Eq.~(\ref{eq:NNLO_meth}) and the
theoretical uncertainty Eq.~(\ref{eq:NNLO_MHOU}) we get
\begin{equation}
\as^\text{NNLO}(m_Z) = 0.1185 \pm 0.0005^{\rm exp}\pm 0.0001^{\rm meth}\pm0.0011^{\rm th}=
0.1185 \pm 0.0012~(1\%)\,,
\label{finvalue}
\end{equation}
where in the last step we have added all uncertainties in quadrature.
For a comparison to other determinations, such as the PDG average, we
recommend using only 
the experimental uncertainty (the methodological
uncertainty being negligible), which reflects the limitations of our
result and procedure, but not the limitation due to the fact that our
result is obtained at NNLO. For precision phenomenology, however, we
recommend use of the total uncertainty in order to conservatively
account for the MHOU.

This result can be compared to the previous
one~\cite{Ball:2011us} based on NNPDF2.1,  $\as^\text{NNLO}(m_Z) = 0.1173 \pm
0.0007^\text{exp}\pm 0.0009^\text{th}$. In comparison to this older
result,  
the central value 
of $\as(m_Z)$ has increased
by $\Delta \alpha_s=+0.0012$ .
 As far as uncertainties are concerned,
both the theoretical and experimental uncertainties on this previous
result are larger, if one compares like with like.
The experimental
uncertainty should actually be compared to Eq.~(\ref{eq:globalnnlo_central})
as it was obtained with the same method. The uncertainty is somewhat underestimated
 because it neglects the correlation between PDFs and
$\alpha_s$, while the theory uncertainty should be compared to
 Eq.~(\ref{eq:MHOUnn}) which is also based on the CH method.
We conclude that, in comparison to Ref.~\cite{Ball:2011us}, 
the current result is more precise, though with
more conservatively estimated uncertainties.

\begin{figure}[t]
\begin{center}
  \includegraphics[scale=1.25]{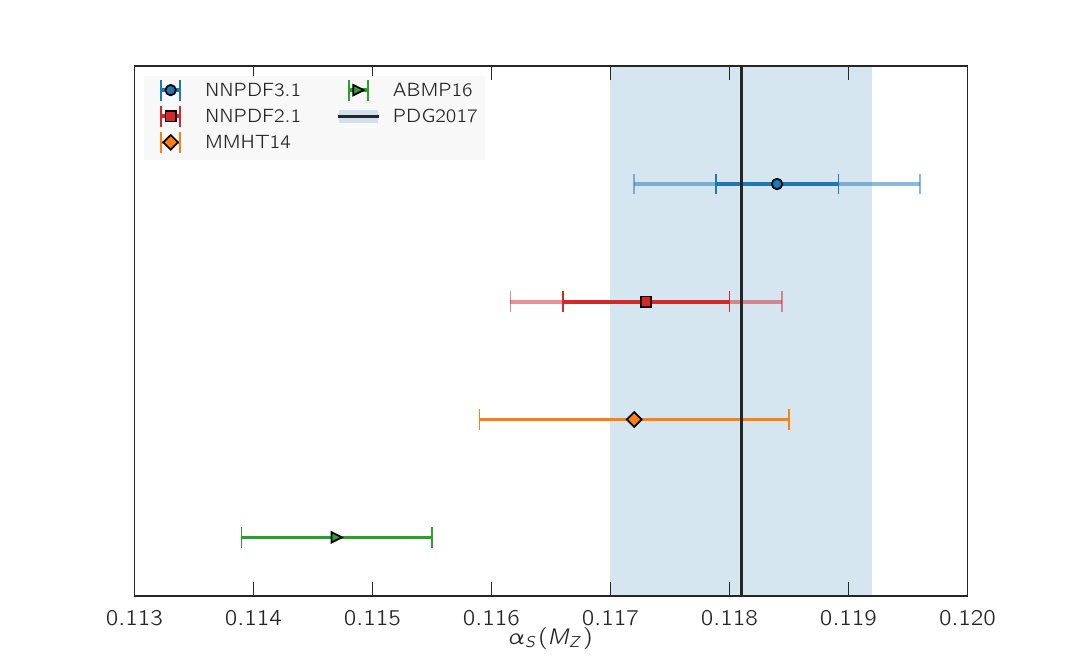}
  \caption{\small Comparison of the present NNLO determination of
    $\amz$, Eq.~(\ref{finvalue}), with the PDG average and with the previous 
  ABMP16, MMHT14, and NNPDF2.1 results.
  For the NNPDF values, the inner (darker) error bar correspond to
   experimental uncertainties, while the outer (lighter) one
  indicates the sum in quadrature of experimental and theoretical
  uncertainties.
  }
\label{fig:alphas_comparison}
\end{center}
\end{figure}

In Fig.~\ref{fig:alphas_comparison} we
compare the NNLO result of Eq.~(\ref{finvalue}) to our
previous result~\cite{Ball:2011us}, to the current PDG average~\cite{Patrignani:2016xqp},
and to two recent determinations obtained from simultaneous fit of
PDFs and $\amz$,  ABMP16~\cite{Alekhin:2017kpj}
and  MMHT2014~\cite{Harland-Lang:2015nxa}.
We find good agreement with the
PDG average as well as with the
MMHT14 and NNPDF2.1 determinations.
It has been suggested~\cite{Ball:2013gsa,Thorne:2014toa} that the
lower ABMP16 value can be partly explained by the use of a
fixed-flavour
number scheme with  $N_f=3$ for the treatment of DIS
data.
It is 
interesting to observe that the current AMBP16 value is higher than  
previous values of $\amz$
obtained by the same group~\cite{Alekhin:2013nda}, from which the
ABMP16 analysis in particular 
differs  because of inclusion in Ref.~\cite{Alekhin:2017kpj} 
of LHC top  production and $W$ and $Z$ production data (described with $N_f=5$).

Interestingly, the $\amz$ determination from the NNPDF3.1 fit is higher than any other recent
determination from PDF fits.
Inspection of
Figs.~\ref{fig:as_bestfit_replicas} and~\ref{fig:cumulative_chi2_diff}
strongly suggests that this increase is driven by the high-precision
LHC data, especially for gauge boson production (including the $Z$
$p_T$ distribution) but also for top and jet production.

%%%%%%%%%%%%%%%%%%%%%%%%%%%%%%%%%%%%%%%%%%%%%%%%%%%%%%%%%%%%%%%%%%%%%
%%%%%%%%%%%%%%%%%%%%%%%%%%%%%%%%%%%%%%%%%%%%%%%%%%%%%%%%%%%%%%%%%%%%%%

\section{Summary and outlook}
\label{sec:conclusion}

In this work we have presented a new determination of the strong coupling constant $\amz$ jointly with a global
determination of PDFs which, by relying on NNPDF3.1, for the first
time includes a large amount of LHC data using exact NNLO theory in all cases.
In comparison to a previous
determination based on NNPDF2.1, our results exploit the new
correlated replica method that is equivalent to the simultaneous fit of
PDFs and $\as$.
This new method thus fully accounts for the correlations between PDFs and $\as$
in the determination of the best-fit value of $\as$ and of the associated uncertainty.

We find that the determination of $\amz$ is considerably stabilized by
the use of a wide set of different processes and data, and we provide
evidence that a global simultaneous determination of $\amz$ and PDFs
leads to a more stable and accurate result than the one obtained from subsets of
data.
We thus obtain a value of $\amz$ which is likely to be more
precise and more accurate than previous results based on similar
techniques.
We find that the LHC data consistently lead to an increase in the
central value of $\amz$, and observe good overall consistency between the datasets
entering the global fit.
Our NNLO determination turns out to be in agreement
within uncertainties with previous results from global fits  and with the PDG average.

The main limitation of our result comes from the  lack of a reliable method
to estimate the uncertainties related to missing higher
order perturbative corrections.
Theoretical progress in this direction is needed, and perhaps expected, and would be
a major source of future improvement.
For the time being, even with
a very conservative estimate of the theoretical uncertainty, our result
provides one of the most accurate determinations of $\amz$ available,
and thus provides valuable input for precision tests of the Standard
Model and for searches for new physics beyond it.

\subsection*{Acknowledgments}

E.~S. and J.~R. are supported by an European Research Council Starting
Grant ``PDF4BSM''. J.~R. is also supposed by the Netherlands Organization for Scientific
Research (NWO).
The work of Z.K. is supported by the European Research Council
Consolidator Grant ``NNLOforLHC2'' and by the
Executive Research Agency (REA) of the European Commission under the
Grant Agreement
PITN-GA-2012-316704  (HiggsTools).
S.~C. is supported by the HICCUP ERC Consolidator grant (614577).
S.~C. and S.~F. are supported by the
European Research Council under the European Union's Horizon 2020 research and
innovation Programme (grant agreement n$^{\circ}$ 740006). R.~D.~B and L.~D.~D. are supported by UK STFC grants ST/L000458/1 and ST/P0000630/1.

%%%%%%%%%%%%%%%%%%%%%%%%%%%%%%%%%%%%%%%%%%%%%%%%%%%%%%%%%%%%%%%%%%%%%%%%%%%%5

\bibliography{nnpdf31alphas}

\end{document}